\documentclass[12pt]{article}

\usepackage[margin=1.0in]{geometry}
 
\usepackage{graphicx, subcaption}      
\usepackage{amssymb, amsmath}
\usepackage{tikz}
\usetikzlibrary{patterns,angles,quotes,arrows}
\usepackage[shortlabels]{enumitem}
\usepackage{amsthm}
\newtheoremstyle{mystyle}%    % Name
  {}%                         % Space above
  {}%                         % Space below
  {}%                         % Body font
  {}%                         % Indent amount
  {\bfseries}%                % Theorem head font
  {.}%                        % Punctuation after theorem head
  { }%                        % Space after theorem head, ' ', or \newline
  {}%                         % Theorem head spec (can be left empty, meaning `normal')

\theoremstyle{mystyle}
\newtheorem{thm}{Theorem}
\newtheorem{problem}{Problem}
\newtheorem{mydef}{Definition}
\newtheorem{prop}{Proposition}
\newtheorem*{pf}{Proof}
\newtheorem{pf_numbered}{Proof}
\newtheorem*{remark}{Remark}
\newtheorem*{cor}{Corollary}
\usepackage[numbers]{natbib}

\providecommand{\keywords}[1]{\textbf{\textit{Index terms---}} #1}
\usepackage{cite}
\usepackage{appendix}

\begin{document}

\title{Designing Resilient Linear Driftless Systems}

\author{Jean-Baptiste~Bouvier and Melkior Ornik
\thanks{Jean-Baptiste Bouvier and Melkior Ornik are with the Department of Aerospace Engineering and the Coordinated Science Laboratory, University of Illinois at Urbana-Champaign, Urbana, IL 61801, USA.\hfill \break
e-mail: bouvier3@illinois.edu \& mornik@illinois.edu \hfill \break
This work was supported by an Early Stage Innovations grant from NASA’s Space Technology Research Grants Program, grant no. 80NSSC19K0209.}}

\date{}

\maketitle

\begin{abstract}
Critical systems must be designed resilient to all kinds of malfunctions. We are especially interested by the loss of control authority over actuators.
This malfunction considers actuators producing uncontrolled and possibly undesirable inputs.
We investigate the design of resilient linear systems capable of reaching their target even after such a malfunction.
In contrast with the settings of robust control and fault-tolerant control, we consider undesirable but observable inputs of the same magnitude as controls since they are produced by a faulty actuator of the system. The control inputs can then depend on these undesirable inputs.
Building on our previous work, we focus on designing resilient systems able to withstand the loss of one or multiple actuators. Since resilience refers to the existence of a control law driving the state to the target, we naturally continue with the synthesis of such a control law. We conclude with a numerical application of our theory on the ADMIRE fighter jet model.
\end{abstract}

\keywords{Linear systems, Reachability analysis, Control design, Reliability, \break
\hspace*{38mm} Redundancy.}

%===============================================================================

\section{Introduction}

Redundancy is the key to guarantee the resilience of a system, as proven by NASA during the space race \citep{NASA_redundancy}. We focus on the resilience of linear systems to the loss of control authority over some of their actuators. This malfunction studied in \citep{Melkior, IFAC} refers to actuators producing uncontrolled and possibly undesirable outputs. Thanks to sensors on each actuators and a fault-detection mechanism as in \citep{actuators_measures}, the controller has real-time readings over all inputs even the uncontrolled ones and can identify defective actuators.

This paper is a continuation of our initial work in \citep{IFAC} and investigates how to design linear systems resilient to a loss of control authority over some of their actuators, i.e., that can still reach their initial target.
We say that a target is \textit{resiliently reachable} from an initial state if for any undesirable inputs, there exists a control law --- possibly dependent on current undesirable inputs, but with no knowledge of future ones --- able to drive the system to the target. 
While not referring to it as resilient reachability, \citep{Trieste} and \citep{Ian_Mitchell} considered this setting but developed complex algorithms giving absolutely no hindsight about how to design resilient systems or how to synthesize a resilient control input, which are our two objectives. Moreover, the resilience analysis of a system, like the one performed in Section~\ref{subsec:Admire} is impossible with the methods of \citep{Trieste, Ian_Mitchell}.
Building on \citep{Delfour}, the work \citep{IFAC} established reachability conditions for linear systems, but it did not investigate resilience of systems.

Loss of control authority over actuators is not covered by fault-tolerant papers as they consider either actuators locking in place \citep{actuator_lock}, losing effectiveness but remaining controllable \citep{fault_tolerance}, or a combination of both \citep{hypersonic_fault_tolerant}, but not uncontrolled and fully effective actuators.
While the field of robust control \citep{Bertsekas, kurzhanski2002} encompasses our type of malfunction, it is too conservative to solve our problem. Indeed, our undesirable inputs can have the same magnitude as the controlled inputs and thus are too large to be handled by a robust control law \citep{limits_of_robust}.
Moreover, the robust control setting treats undesirable inputs as unknown, while we assume to have real-time readings of them. Thus, our resilient controller adapts to the undesirable inputs and performs much better than an overly conservative robust controller, as demonstrated in Section~\ref{subsec:robust}.

Our objective is to design linear systems resilient to the loss of control authority over some of their actuators with a minimal redundancy. The contributions of this paper are twofold.
First, we determine the minimal degree of overactuation necessary to design a resilient system. 
Second, we synthesize a control law driving a resilient system's state to its target despite loss of control authority over some actuators.
To establish these results, we will first focus on driftless systems, a common application in robotics \citep{robotics}, before extending our findings to systems with drift.

The remainder of the paper is organized as follows.
Section~\ref{sec:def} defines the problems of interest and introduces the preliminary results from \citep{IFAC}.
In Section~\ref{sec:resilient_matrix}, we develop the notion of resilient control matrices and we determine their minimal size in Section~\ref{sec:min_size}.
Building on the driftless case, Section~\ref{sec:synthesis} focuses on the synthesis of a resilient control law for linear systems with and without drift.
We illustrate our theory in Section~\ref{sec:examples} with three scenarios featuring a model of a fighter jet undergoing a loss of control authority.

\vspace{2mm}
\textit{Notation:} The identity matrix of size $n$ is denoted $I_n$. The transpose of a matrix $M$ is $M^\top$, a positive semidefinite matrix is denoted by $M \succeq 0$ and a positive definite matrix by $M \succ 0$. The eigenvalues of a square matrix $M$ are gathered in $\lambda(M) := \{z: \det(zI_n - M) = 0\}$. The singular values of a matrix $M$ are the $\sigma^M \geq 0$ such that $\det\big( (\sigma^M)^2 I_n - M^\top M \big) = 0$, and $\sigma_{max}^M = \max \sqrt{\lambda(M^\top M)}$.

The vector $e_i$ is composed of zeros except its $i^{th}$ element is one. The column vector $z = (z_1,\hdots, z_n) \in \mathbb{R}^n$ has a norm $\|z\| = \sqrt{\sum z_i^2}$. We use $\langle \cdot, \cdot \rangle$ to denote the inner product between vectors.
 
The set of integers from $1$ to $n$ is denoted by $[n]$. The unit sphere in $\mathbb{R}^n$ is denoted by $\mathbb{S} = \big\{ x \in \mathbb{R}^n : \|x\| = 1\big\}$ , while $\mathbb{B}_X(c, \varepsilon) = \big\{x \in X : \|x-c\| \leq \varepsilon \big\}$ is the ball of center $c$ and radius $\varepsilon$ in the space $X$. The ellipsoid of center $c$ and shape matrix $P \succ 0$ is $\mathcal{E}(c, P) = \big\{x : (x - c)^\top P (x - c) \leq 1\big\}$.

The space of square integrable functions $u:[0, T] \rightarrow \mathbb{R}^m$ is denoted by $\mathcal{L}_2\big([0, T], \; \mathbb{R}^m \big)$ or simply $\mathcal{L}_2$, and contains all functions with a finite $\mathcal{L}_2$-norm: $\|u\|^2 = \int_0^T \|u(t)\|^2 dt$.

Operators $Tr$ and $\det$ respectively denote the trace and the determinant of a matrix. To obtain the real part of a complex number we use the operator $Re:\mathbb{C} \rightarrow \mathbb{R}$.
The quantifiers $\exists$ and $\forall$ denote ``there exists'' and ``for all'', respectively.
For $p \leq m \in \mathbb{N}$, we denote the number of $p$-combinations among $m$ elements with the binomial coefficient $\Big(\hspace{-2mm}\def\arraystretch{0.7}\begin{array}{c} \footnotesize{m} \\ \footnotesize{p} \end{array}\hspace{-2mm}\Big)$.

\section{Problem Statement}\label{sec:def}

Consider a system governed by the differential equation 
\begin{equation}\label{eq:initial_ODE}
    \dot{x} = Ax + \bar{B} \bar{u}, \qquad x(0) = x_0 \in \mathbb{R}^n,
\end{equation}
where $A \in \mathbb{R}^{n \times n}$ and $\bar{B} \in \mathbb{R}^{n \times m}$ are constant matrices with $n$ and $m \in \mathbb{N}$. 
Assume that the control specification is one of reachability.
In other words, let $G \subset \mathbb{R}^n$ be the \textit{target ball} of radius $\varepsilon \geq 0$ around $x_{goal} \in \mathbb{R}^n$ to be reached by the system.
Assume that during its mission the system loses control authority over $p$ of its $m$ actuators, with $p < m$. 
These $p$ actuators are then producing uncontrolled and possibly undesirable inputs.
Thanks to a fault-detection mechanism relying on sensors on each actuators, we can separate the controlled inputs $u \in \mathbb{R}^{m-p}$ from the undesirable inputs $w \in \mathbb{R}^p$ by writing $\bar{u} = (u^\top,\ w^\top)$ and $\bar{B} = [B\ C]$, with $B \in \mathbb{R}^{n \times (m-p)}$ and $C \in \mathbb{R}^{n \times p}$. The system's dynamics can thus be rewritten as follows:
\begin{equation}\label{eq:ODE}
    \dot{x}(t) = A x(t) + B u(t) + C w(t), \qquad x(0) = x_0 \in \mathbb{R}^n.
\end{equation}

The technical work of this paper follows the assumptions of \citep{IFAC, Delfour} by considering inputs of finite energy, that are thus square integrable signals. Namely, if $U$ is the set of admissible control laws and $W$ is the set of undesirable inputs, we consider
\begin{align}\label{eq:def of U, W, G}
    U &= \big\{u \in \mathcal{L}_2\big([0, T], \; \mathbb{R}^{m-p} \big) : \|u\| \leq 1 \big\} = \mathbb{B}_{\mathcal{L}_2}(0,1),\nonumber \\
    W &= \big\{w \in \mathcal{L}_2\big([0, T], \; \mathbb{R}^p \big) : \|w\| \leq 1 \big\} = \mathbb{B}_{\mathcal{L}_2}(0,1), \\
    G &= \big\{x \in \mathbb{R}^n : \|x - x_{goal}\| \leq \varepsilon \big\} = \mathbb{B}_{\mathbb{R}^n}(x_{goal},\varepsilon). \nonumber
\end{align}
We want to determine what kind of system is still able to reach its target after a loss of control over some of its actuators.
We can now define a \textit{resilient system} as follows

\begin{mydef}
    The system $(A, \bar{B})$ following the dynamics \eqref{eq:initial_ODE} is \textit{resilient} to the loss of control authority over the $p$ actuators represented by $C$ if for any target ball $G$ and any undesirable input $w \in W$, there exists a control law $u_w \in U$ driving the state of \eqref{eq:ODE} from $x_0$ to $G$.
\end{mydef}

We note that, as in \citep{IFAC}, the control law $u_w$ can depend on the undesirable input $w$. Unlike the concept of strong reachability in classical robust control \citep{Bertsekas, kurzhanski2002, Tubes, Robust}, the objective is not to a priori design a control law working for any perturbation, but instead to have a control law for each undesirable input.
Indeed, we assumed to have sensors on each actuators so that all inputs to the system are available to the controller.
Therefore, resilient reachability guarantees that whatever the undesirable inputs are, there is a control law \textit{dependent on the undesirable inputs} driving the system to its target.
The intuitive expectation behind this dependency is that such a controller can handle undesirable inputs of a larger magnitude than a standard robust controller.

Since a resilient system can operate with fewer actuators than in its nominal configuration, we have the intuition that such a system must be initially \textit{overactuated}.

\begin{mydef}
    A system is \textit{overactuated} if the control matrix $\bar{B}$ has strictly more columns than rows.
\end{mydef}
We can now formulate our two main objectives.

\begin{problem}\label{prob:minimal size}
    Determine the minimal degree of overactuation required to build a resilient system.
\end{problem}

Since the definition of a resilient system calls for the existence of a control law, we are naturally led to our second objective.

\begin{problem}\label{prob:control synthesis}
    For a resilient system sustaining an undesirable input $w$, synthesize a control law $u_w$ that drives the system's state $x(t)$ to the target $G$.
\end{problem}

The resilience of a linear system \eqref{eq:initial_ODE} is mostly determined by its control  matrix $\bar{B}$. Therefore, in the next two sections we first focus on driftless systems, i.e., where \eqref{eq:ODE} becomes
\begin{equation}\label{eq:driftless_ODE}
    \dot{x}(t) = B u(t) + C w(t), \qquad x(0) = x_0 \in \mathbb{R}^n.
\end{equation}

These systems have been studied in \citep{IFAC} where we established conditions to verify whether a target is \textit{resilient reachability at a certain time}.

\begin{mydef}
    The target $G$ is \textit{resiliently reachable at time $T$} from $x_0$ if for any undesirable inputs $w \in W$, there exists a control law $u_w \in U$ that drives the system following \eqref{eq:ODE} to $x(T) \in G$.
\end{mydef}

In the case where the matrix $B$ is invertible the problem of resilient reachability becomes trivial. Indeed, the control law $u_w = -B^{-1}Cw$ would completely counteract the undesirable inputs.
However, we are interested in general matrices $B$.
For those systems, the work in \citep{IFAC} offers a straightforward expression to evaluate reachability at a certain time.

\begin{thm}\label{thm:reach}
$G $ is resiliently reachable at time $T$ from $x_0$ if and only if
$$\underset{h\, \in\, \mathbb{S}}{\max}\Big\{ \langle h,x_0 - x_{goal} \rangle - \sqrt{T} \left\|B^\top h\right\| + \sqrt{T} \left\|C^\top h\right\| \Big\} \leq \varepsilon.$$
\end{thm}

The condition in Theorem~\ref{thm:reach} is simplified in \citep{IFAC} with the definitions of $d = x_{goal} - x_0$ and of the function
\begin{equation}\label{eq:function_g}
    g(h) := \left\|C^\top h\right\| - \left\|B^\top h\right\| \qquad \text{for}\ h \in \mathbb{S}.
\end{equation}
Theorem~\ref{thm:reach} only states whether $G$ is reached exactly at $T$. The situations where the target must instead be reached before a time limit, call for \textit{resilient reachability by time $T$}.

\begin{mydef}
    The target $G$ is \textit{resiliently reachable by time $T$} if there exists a time $t \leq T$ at which $G$ is resiliently reachable.
\end{mydef}

Then, \citep{IFAC} described reachability by time $T$ as a minimax problem. The target $G$ is resiliently reachable from $x_0$ by time $T$ if and only if
$\underset{t\, \in\, [0,T]}{\min}\Big\{ \underset{h\, \in\, \mathbb{S}}{\max}\big\{ h^\top d + g(h) \sqrt{t} \big\} \Big\} \leq \varepsilon.$

\begin{thm}\label{thm:sign of g}
The following statements hold:
\begin{enumerate}[(a), parsep=0mm, topsep=0mm]
   \item If $\underset{h\, \in\, \mathbb{S}}{\max}\big\{g(h)\big\} < 0$, there exists a time $t_{lim}$ such that $G$ is resiliently reachable at time $t$ for all $t \geq t_{lim}$.
   \item If $\underset{h\, \in\, \mathbb{S}}{\max}\big\{g(h)\big\} > 0$, there exists a time $t_{lim}$ such that $G$ is not resiliently reachable at time $t$ for all $t > t_{lim}$.
  \item If $\underset{h\, \in\, \mathbb{S}}{\max}\big\{g(h)\big\} = 0$, the resilient reachability of $G$ depends on the distance $d$.
\end{enumerate}
\end{thm}

The maximum of $g$ can be difficult to compute, so we establish a more straightforward criteria for resilient reachability that is easily computable.

\begin{thm}\label{thm:positive_def}
For $F := BB^\top - CC^\top$, the following statements hold:
\begin{enumerate}[(a), parsep=0mm, topsep=0mm]
    \item If $F \succ 0$, there exists a time $t_{lim}$ such that $G$ is resiliently reachable at time $t$ for all $t \geq t_{lim}$.
    \item  If $F \nsucceq 0$, there exists a time $t_{lim}$ such that $G$ is not resiliently reachable at time $t$ for all $t > t_{lim}$.
\end{enumerate}
\end{thm}

\begin{pf}
The statement (a) is equivalent to Theorem \ref{thm:sign of g} (a):
\begin{align*}
    \underset{h\, \in\, \mathbb{S}}{\max}\ g(h) < 0 &\Longleftrightarrow\ \forall\ h \in \mathbb{S}, \quad \|C^\top h\| - \|B^\top h\| < 0 \\
    &\Longleftrightarrow\ \forall\ h \in \mathbb{S}, \quad h^\top CC^\top h < h^\top BB^\top h \\
    &\Longleftrightarrow\ \forall\ h \in \mathbb{S}, \quad 0 < h^\top F h \\
    &\Longleftrightarrow\ \forall x \in \mathbb{R}^{n}\backslash\{0\}, \quad 0 < \frac{x^\top F x}{\|x\|^2} \\
    &\Longleftrightarrow \quad F \succ 0.
\end{align*}
And similarly, statement (b) is equivalent to Theorem \ref{thm:sign of g} (b):
\begin{align*}
    \underset{h\, \in\, \mathbb{S}}{\max}\ g(h) > 0 &\Longleftrightarrow\ \exists\ h\ \in\ \mathbb{S}\ :\ \|C^\top h\| - \|B^\top h\| > 0\\
    &\Longleftrightarrow\ \exists\ h\ \in\ \mathbb{S}\ :\ h^\top CC^\top h > h^\top BB^\top h  \\
    &\Longleftrightarrow\ \exists\ h\ \in\ \mathbb{S}\ :\ h^\top F h < 0 \\
    &\Longleftrightarrow\ \quad F \nsucceq 0. \quad \blacksquare
\end{align*}
\end{pf}

We have thus obtained simple analytical conditions concerning the resilient reachability of a target.

\section{Resilient Control Matrices}\label{sec:resilient_matrix}

A driftless system is entirely described by its control matrix $\bar{B}$. Thus, our overarching idea is to link the resilience of a driftless system to the properties of its control matrix.

When losing control authority over $p$ of the $m$ actuators of the system, we remove the corresponding columns $j_1, \hdots, j_p$ from $\bar{B}$ to form the matrix $C$ and we name $B$ the remaining control matrix. 
We can now define a \textit{$p$-resilient control matrix}.

\begin{mydef}\label{def:p_resilience}
    The control matrix $\bar{B} \in \mathbb{R}^{n \times m}$ is \textit{$p$-resilient} if for all pairwise distinct indices $j_1, \hdots, j_p \in [m]$ the system following the driftless dynamics \eqref{eq:driftless_ODE} can resiliently reach any target ball. 
\end{mydef}

The \textit{degree of resilience} of the matrix $\bar{B}$ is the highest $p$ for which $\bar{B}$ is $p$-resilient. Definition~\ref{def:p_resilience} implies that if a control matrix is $p$-resilient, then it is also ($p$-1)-resilient. On the other hand, if a control matrix is not $p$-resilient, then it is not ($p$+1)-resilient either.

\subsection{Necessary and sufficient conditions for $p$-resilience}

Based on our previous work, we derive two necessary and sufficient criteria to verify if a control matrix is resilient.

\begin{prop}\label{prop:B_bar_g<0}
     The control matrix $\bar{B} \in \mathbb{R}^{n \times m}$ is $p$-resilient if and only if $\underset{h\, \in\, \mathbb{S}}{\max}\ g(h) < 0$ for all pairwise distinct $j_1, \hdots, j_p \in [m]$, with $g(h) = \|C^\top h\| - \|B^\top h\|$.
\end{prop}
\begin{pf}
If $\underset{h\, \in\, \mathbb{S}}{\max}\ g(h) < 0$ for all pairwise distinct indices $j_1, \hdots, j_p \in [m]$, then from Theorem \ref{thm:sign of g}, any target ball is resiliently reachable by the system of dynamics \eqref{eq:driftless_ODE}, so $\bar{B}$ is $p$-resilient.

On the other hand, assume that $\bar{B}$ is $p$-resilient. For all pairwise distinct $j_1, \hdots, j_p \in [m]$, the continuous function $g$ reaches a maximum $g_{max}$ over the compact set $\mathbb{S}$. If $g_{max} > 0$, then from Theorem \ref{thm:sign of g} (b) after some time, any target ball becomes not resiliently reachable, which contradicts the resilience of $\bar{B}$. If $g_{max} = 0$, then from Theorem \ref{thm:sign of g} (c) there are some balls that are not resiliently reachable. It also contradicts the resilience of $\bar{B}$. Therefore, $g_{max} < 0. \quad \blacksquare$
\end{pf}

Computing the maximum of each function $g$ can be difficult. Thus, we employ Theorem \ref{thm:positive_def} to simplify the result of Proposition \ref{prop:B_bar_g<0}. As previously, we create $C$ by removing $p$ columns from $\bar{B}$, indexed by $j_1,\hdots,j_p$ and call $B$ the remaining control matrix.

\begin{prop}\label{prop:B_bar_F_pos_def}
    The matrix $\bar{B}$ is $p$-resilient if and only if $F = B B^\top - C C^\top \succ 0$ for all pairwise distinct $j_1, \hdots, j_p \in [m]$.
\end{prop}
\begin{pf}
The result follows directly from Proposition \ref{prop:B_bar_g<0} and the proof of Theorem \ref{thm:positive_def}. $\blacksquare$
\end{pf}

Proposition \ref{prop:B_bar_F_pos_def} enables us to determine $p$-resilience of a system with $m$ actuators by verifying the positive definiteness of $\Big(\hspace{-2mm}\def\arraystretch{0.7}\begin{array}{c} \footnotesize{m} \\ \footnotesize{p} \end{array}\hspace{-2mm}\Big)$ matrices.
Before proceeding further, we need to establish a less obvious necessary condition for 1-resilience.

\begin{prop}\label{prop:B_bar_invertible}
    If $\bar{B}$ is 1-resilient, then $\bar{B}\bar{B}^\top \succ 0$.
\end{prop}
\begin{pf}
    Assume that $\bar{B}\bar{B}^\top$ is not positive definite. Then, there exists $x~\neq~0$ such that $x \bar{B}\bar{B}^\top x \leq 0$.
    Without loss of generality, assume we remove the last column $C$ from $\bar{B}$:
    $$\bar{B}\bar{B}^\top = \begin{bmatrix} B & C\end{bmatrix} \begin{bmatrix} B^\top \\ C^\top \end{bmatrix} = B B^\top + C C^\top.$$
    So $F = BB^\top - CC^\top = \bar{B}\bar{B}^\top - 2CC^\top$.\hfill \break
    Then $x^\top F x = x^\top \bar{B}\bar{B}^\top x - 2 x^\top CC^\top x \leq 0 - 2\|C^\top x \|^2 \leq 0$, so $F$ is not positive semidefinite. By Proposition \ref{prop:B_bar_F_pos_def}, $\bar{B}$ is not 1-resilient. $\quad \blacksquare$
\end{pf}

With these results, we can start to formalize our initial intuition about overactuation.

\begin{prop}\label{prop:overactuation}
    If $\bar{B}$ is 1-resilient, then the system is overactuated.
\end{prop}
\begin{pf}
    Assume $\bar{B} \in \mathbb{R}^{n \times m}$ is not overactuated, then $m~\leq~n$.
    After losing control of one actuator, the remaining control matrix $B$ has $n$ rows and at most $n-1$ columns.
    From \citep{matrices}, the rank of a matrix is smaller than its smallest dimension, so $rank(B) \leq n-1$. The rank of a product of matrices is smaller than the rank of each of the matrices \citep{matrices}, so $rank(BB^\top)~\leq~rank(B)$.
    Thus, $rank(BB^\top) \leq n-1$.
    
    Since $BB^\top$ is a square matrix of size $n$, it is not invertible. Then, $BB^\top$ is not positive definite, so $F = BB^\top - CC^\top$ is not positive definite either. According to Proposition \ref{prop:B_bar_F_pos_def}, $\bar{B}$ is not 1-resilient. $\quad \blacksquare$
\end{pf}

It is intuitive that a system without redundancy among actuators cannot be resilient, because a malfunctioning actuator cannot be counteracted.
On the other hand, if there are many copies of each actuator, then the system can lose control of one and still be functioning.
In between these extremes there is a minimum degree of overactuation required for resilience. Since adding actuators in practice comes with a cost, determining the minimal size of a resilient matrix can help reducing that cost.

\subsection{Resilience invariant and Singular Value Decomposition}

The degree of resilience of a matrix is left unchanged when applying some basic transformations. Determining those will help our study of the minimal size of a resilient matrix.

\begin{prop}\label{prop:resil_inv}
    The degree of resilience is not affected by left multiplication by an invertible matrix. 
\end{prop}
\begin{pf}
    Let $\bar{B}$ be a $p$-resilient control matrix, and $P$ an invertible matrix of adequate size. The modified control matrix is $\bar{B}^P = P\bar{B}$. We extract $p$ columns of $\bar{B}^P$ to create $B^P = PB$ and $C^P = PC$. Then,
    \begin{align*}
        F^P &= B^P \big(B^P\big)^\top - C^P \big(C^P\big)^\top \\
        &= \big(PB\big) \big(PB\big)^\top - \big(PC\big) \big(PC\big)^\top \\
        &= PB B^\top P^\top - PC C^\top P^\top = P F P^\top,
    \end{align*}
    with $F = B B^\top - C C^\top$. 
    Because $P$ is invertible, we know from \citep{matrix_computations} that $F \succ 0$ if and only if $F^P \succ 0$. Using Proposition \ref{prop:B_bar_F_pos_def}, we conclude that $\bar{B}^P$ is also $p$-resilient. $\quad \blacksquare$
\end{pf}

We note that rotations, permutations of columns and non-zero scaling are all invertible operations, and thus do not change the degree of resilience of a matrix.
We can now simplify the resilience investigation with the Singular Value Decomposition (SVD).

Let $\bar{B} \in \mathbb{R}^{n \times m}$. The compact SVD \citep{downdating} of $\bar{B}$ is $UDV$, with $U$ orthogonal of size $n \times n$, $D$ a diagonal matrix gathering the $n$ singular values of $\bar{B}$, and $V$ of size $n \times m$ with orthonormal rows: $VV^\top = I$. 

\begin{prop}\label{prop:SVD}
    The following statements hold for $p \geq 1$:
    \begin{enumerate}[(a), itemsep=0mm, topsep=0mm]
        \item If $\bar{B}$ is $p$-resilient, then $V$ is also $p$-resilient.
        \item If $V$ is $p$-resilient and $\bar{B}\bar{B}^\top \succ 0$, then $\bar{B}$ is $p$-resilient.
    \end{enumerate}
\end{prop}
\begin{pf}
    For statement (a), assume that $\bar{B}$ is $p$-resilient with $p \geq 1$. Then, Proposition \ref{prop:B_bar_invertible} states that $\bar{B}\bar{B}^\top \succ 0$. Thus, the singular values of $\bar{B}$ are non-zero \citep{downdating}. Then, the diagonal matrix $D$ is invertible. The matrix $U$ is orthogonal so it is also invertible. Therefore, $\bar{B} = UDV$ and $V$ have the same degree of resilience according to Proposition \ref{prop:resil_inv}.
    
    \vspace{2mm}
    For statement (b), since $\bar{B}\bar{B}^\top \succ 0$, the matrix $D$ is invertible. Then, by Proposition \ref{prop:resil_inv} the matrix $\bar{B}$ has the same degree of resilience as $V$.
    $\quad \blacksquare$
\end{pf}

Since $V$ has orthonormal rows, we proceed to study the $p$-resilience of $V$ instead of $\bar{B}$. 
Let $C_V$ be any matrix formed with $p$ columns taken from $V$, and $B_V$ the associated remaining control matrix.

\begin{prop}\label{prop:max sing val p-resilience}
    The matrix $V \in \mathbb{R}^{n \times m}$ with orthonormal rows is $p$-resilient if and only if $\sigma_{max}^{C_V^\top} < \frac{1}{\sqrt{2}}$ for all $\Big(\hspace{-2mm}\def\arraystretch{0.7}\begin{array}{c} \footnotesize{m} \\ \footnotesize{p} \end{array}\hspace{-2mm}\Big)$ possible $C_V$ matrices.
\end{prop}
\begin{pf}
    We extract $p$ columns from $V$ to create $B_V$ and $C_V$ and we investigate whether $F_V := B_V B_V^\top - C_V C_V^\top$ is positive definite. Without loss of generality, $V = [ B_V\ C_V ]$, so that
    $VV^\top = B_V B_V^\top + C_V C_V^\top$. The matrix $V$ has orthonormal rows: $VV^\top \hspace{-1mm} = I_n$. Then, $F_V = \hspace{-0.5mm} VV^\top \hspace{-1mm} - \hspace{-0.5mm} 2C_V C_V^\top \hspace{-0.5mm} = \hspace{-0.5mm} I_n \hspace{-1mm} - \hspace{-0.5mm} 2C_V C_V^\top$.
    Let $\lambda$ be an eigenvalue of $F_V$. Then,
    \begin{align*}
        0 &= \det\big( \lambda I_n - F_V\big) = \det\big( \lambda I_n - I_n + 2C_V C_V^\top \big) \\
        &= \det\big( (\lambda-1) I_n + 2C_V C_V^\top \big) = \big(-2\big)^n \det\Bigg( \Big(\frac{1-\lambda}{2}\Big) I_n - C_V C_V^\top \Bigg).
    \end{align*}
    Let us define $s := \frac{1-\lambda}{2}$, so that $s$ is an eigenvalue of $C_V C_V^\top$. Let $x \neq 0$ be an eigenvector such that $C_V C_V^\top x = sx$. A left multiplication by $x^\top$ lead to $\|C_V^\top x\|^2 = s\|x\|^2$, so $s \geq 0$.
    
    Then, $\sqrt{s}$ is a singular value of $C_V^\top$. 
    We note that $\lambda > 0$ if and only if $\sqrt{s} < \frac{1}{\sqrt{2}}$. Since $\sigma_{max}^{C_V^\top}$ is the maximal singular value of $C_V^\top$, $F_V \succ 0$ if and only if $\sigma_{max}^{C_V^\top} < \frac{1}{\sqrt{2}}$.
    $\quad \blacksquare$
\end{pf}

Propositions \ref{prop:SVD} and \ref{prop:max sing val p-resilience} greatly simplify the investigation of the minimal size of resilient matrices.

\section{Minimal size of resilient matrices}\label{sec:min_size}

\subsection{1-resilient matrices}

We will now establish a necessary condition determining the minimal size of a 1-resilient control matrix.
\begin{thm}\label{thm:min_size}
If $\bar{B} \in \mathbb{R}^{n \times m}$ is 1-resilient, then $m \geq 2n+1$.
\end{thm}
\begin{pf_numbered}
    Let $\bar{B} \in \mathbb{R}^{n \times m}$ be 1-resilient. We extract the column $i \in [m]$ from $\bar{B}$ to form $C_i$, while the remaining control matrix is called $B_i$. \hfill \break
    We showed that $F_i = B_i B_i^\top - C_i C_i^\top = \bar{B}\bar{B}^\top - 2C_i C_i^\top$ in the proof of Proposition \ref{prop:B_bar_invertible}. Therefore,
    \begin{equation*}
        \det(F_i) = \det\big(\bar{B}\bar{B}^\top - 2C_i C_i^\top\big).
    \end{equation*}
    We now employ the matrix determinant lemma \citep{matrices}:
    \begin{equation*}
        \det\big(\bar{B}\bar{B}^\top - 2C_i C_i^\top\big) = \big(1 -2 C_i^\top \big(\bar{B}\bar{B}^\top\big)^{-1} C_i \big)\det\big(\bar{B}\bar{B}^\top\big).
    \end{equation*}
    We sum the previous equations over $i \in [m]$ to obtain
    \begin{equation*}
        \sum_{i=1}^m \det(F_i) = \det\big(\bar{B}\bar{B}^\top\big) \left( m - 2\sum_{i=1}^m C_i^\top \big(\bar{B}\bar{B}^\top\big)^{-1} C_i \right).
    \end{equation*}
    Now, note that
    \begin{align*}
        \sum_{i = 1}^{m} C_i^\top \big(\bar{B}\bar{B}^\top\big)^{-1} C_i &= \sum_{i = 1}^{m} (\bar{B}e_i)^\top \big(\bar{B}\bar{B}^\top\big)^{-1} \bar{B}e_i = \sum_{i = 1}^{m} e_i^\top \bar{B}^\top \big(\bar{B}\bar{B}^\top\big)^{-1} \bar{B} e_i \\
        & = Tr \left( \bar{B}^\top \big(\bar{B}\bar{B}^\top\big)^{-1} \bar{B} \right) = Tr \left( \bar{B} \bar{B}^\top \big(\bar{B}\bar{B}^\top\big)^{-1} \right) = Tr(I_n) = n.
    \end{align*}
    Therefore,
    \begin{equation}\label{eq:size}
        \sum_{i = 1}^{m} \det(F_i) = \det\big(\bar{B}\bar{B}^\top\big)(m - 2n).
    \end{equation}
    Following Proposition \ref{prop:B_bar_invertible}, we know that $\bar{B}\bar{B}^\top \succ 0$, so its determinant is positive. According to Proposition \ref{prop:B_bar_F_pos_def}, we also have that for all $i \in [m]$, $\det(F_i) > 0$. Thus $m - 2n > 0$, i.e., $m \geq 2n+1$. $\quad \blacksquare$
\end{pf_numbered}

We also present an alternate proof of this theorem making use of Propositions \ref{prop:SVD} and \ref{prop:max sing val p-resilience}.

\begin{pf_numbered}
    Similarly as in Proposition \ref{prop:SVD}, we employ the compact SVD on $\bar{B} = UDV$. From the part (a) of Propositon \ref{prop:SVD} the matrix $V \in \mathbb{R}^{n \times m}$ is 1-resilient. The columns of $V$ are denoted $C_j$ and its orthonormal rows $r_i$. Then,
    \begin{equation}\label{eq:sum_columns}
        \sum_{j=1}^m \|C_j\|^2 = \sum_{j=1}^m \sum_{i=1}^n V_{ij}^2 = \sum_{i=1}^n \sum_{j=1}^m V_{ij}^2 = \sum_{i=1}^n \underbrace{\|r_i\|^2}_{=\ 1} = n.
    \end{equation}
    If $\underset{j}{\max} \|C_j\|^2 < \frac{n}{m}$, then it contradicts \eqref{eq:sum_columns}. 
    From \citep{matrix_computations} we also know that the maximum singular value of a column vector is its norm.
    We combine these results with the condition of Proposition \ref{prop:max sing val p-resilience}:
    $$\frac{n}{m} \leq \underset{j}{\max} \|C_j\|^2 = \big(\sigma_{max}^{C^\top}\big)^2 < \frac{1}{2},$$
    so $m \geq 2n + 1$ is a necessary condition for $1$-resilience. $\quad \blacksquare$
\end{pf_numbered}

Theorem \ref{thm:min_size} shows that at least $2n + 1$ actuators are required to have a 1-resilient control system in $n$ dimensions.
We will now prove that $n \times (2n + 1)$ is in fact the minimal size of 1-resilient matrices by producing such a matrix for all $n \in \mathbb{N}$.

\begin{prop}\label{prop:B2}
 For any $n \in \mathbb{N}$, the matrix $\bar{B}~=~[I_n\ I_n\ D]$ with the column $D = \frac{1}{\sqrt{n}}[1 \hdots 1]^\top$ is 1-resilient.
\end{prop}

\begin{pf}
    We will use Theorem \ref{thm:sign of g} and calculate $\underset{h\, \in\, \mathbb{S}}{\max}\ g(h)$ for the loss of any one actuator. First, assume we lose control of one of the first $2n$ columns. Without loss of generality, we assume losing the column $j$ of the first identity matrix, so $C = e_j$.
    Then for $h = (h_1, \hdots, h_n) \in \mathbb{S}$, we have 
    \begin{equation*}
        B^\top h = \left(h_1, \hdots, h_{j-1},\ h_{j+1}, \hdots, h_n,\  h^\top,\ \sum_{i=1}^n \frac{h_i}{\sqrt{n}} \right), 
    \end{equation*}
    \begin{equation*}
        \text{so} \qquad \|B^\top h\|^2 = \sum_{i = 1, \neq j}^n h_i^2 +  \sum_{i = 1}^n h_i^2 + \Bigg(\sum_{i=1}^n \frac{h_i}{\sqrt{n}}\Bigg)^2 = 2\underbrace{\|h\|^2}_{=\ 1} - h_j^2 + \frac{1}{n}\Bigg(\sum_{i=1}^n h_i\Bigg)^2.
    \end{equation*}
    \begin{align}\label{eq:g<0}
        \text{Then,}\qquad g(h) < 0 \quad &\Longleftrightarrow \quad \|C^\top h\|^2 < \|B^\top h\|^2 \quad \Longleftrightarrow \quad h_j^2 < 2 - h_j^2 + \frac{1}{n}\Bigg(\sum_{i=1}^n h_i\Bigg)^2 \nonumber \\
        &\Longleftrightarrow \quad h_j^2 < 1 + \frac{1}{2n}\Bigg(\sum_{i=1}^n h_i\Bigg)^2.
    \end{align}
    If $h_j^2 = 1$, then $h_i = 0$ for all $i \neq j$ because $\|h\| = 1$. Thus $\sum h_i = 1$, so \eqref{eq:g<0} is true. Otherwise, $h_j^2 < 1$ and \eqref{eq:g<0} is also true. Thus, for any $h \in \mathbb{S}$, $g(h) < 0$.
    
    \vspace{1mm}
    The remaining case is when the system loses control of the last actuator. Then $B = [I_n\ I_n]$ and $C = D$. For any $h \in \mathbb{S}$,
    \begin{equation*}
        g(h) = \left|\sum_{i=1}^n\frac{h_i}{\sqrt{n}}\right| - \sqrt{2} \|h\| \leq \frac{\sum |h_i|}{\sqrt{n}} - \sqrt{2}.
    \end{equation*}
    Using the Cauchy-Schwarz inequality \citep{math}, we obtain
    \begin{equation*}
        \sum_{i=1}^n |h_i| \leq \sqrt{\sum_{i=1}^n |h_i|^2}\sqrt{\sum_{i=1}^n 1^2} = \|h\| \sqrt{n} = \sqrt{n}.
    \end{equation*}
    
    Then $g(h) \leq 1 -\sqrt{2} < 0$.
    Therefore, in both cases $\underset{h\, \in\, \mathbb{S}}{\max}\ g(h) < 0$. From Proposition \ref{prop:B_bar_g<0}, the control matrix $\bar{B}$ is 1-resilient. $\blacksquare$
\end{pf}

To sum up, we showed that the minimal size of a 1-resilient control matrix is $n \times (2n + 1)$. 
We will now investigate sufficient conditions allowing to generate 1-resilient control matrices by making use of Proposition \ref{prop:max sing val p-resilience}.

\begin{prop}\label{prop:V_ortho_resilient}
    Any matrix $V \in \mathbb{R}^{n \times m}$ where $m \geq 2n+1$ which has orthonormal rows and whose columns have all the same norm, is 1-resilient.
\end{prop}
\begin{pf}
    Since the columns $C$ of matrix $V$ have the same norm, equation \eqref{eq:sum_columns} implies $\|C\|^2 = \frac{n}{m}$. The maximal singular value of a column vector is its norm \citep{matrix_computations}, so $\sigma_{max}^{C^\top} = \|C\| = \sqrt{\frac{n}{m}}$.
    Since $m \geq 2n+1$, we obtain
    \begin{equation*}
        \frac{n}{m} \leq \frac{1}{2} - \frac{1}{2m} < \frac{1}{2},\quad \text{i.e.},\quad \sigma_{max}^{C^\top} < \frac{1}{\sqrt{2}}. 
    \end{equation*}
    Then, Proposition \ref{prop:max sing val p-resilience} states that $V$ is 1-resilient. $\quad \blacksquare$
\end{pf}

Intuitively, the columns of $V$ having the same norm means that the actuators are equally powerful, whereas the rows having the same norm means that all the states are equally actuated. Furthermore, the orthogonality of rows enforces the necessary condition for \break
1-resilience of Proposition \ref{prop:B_bar_invertible} by making $VV^\top$ positive definite.

With Proposition \ref{prop:V_ortho_resilient} we can now easily generate 1-resilient matrices for any size $n$. For instance,
\begin{equation*}
    \begin{bmatrix} 1 & 1 & 1 & 1 & 1 & 1 \\ 1 & 1 & 1 & -1 & -1 & -1 \end{bmatrix} \quad \text{and} \quad \begin{bmatrix} 1 & 1 & 1 & 1 & 1 & 1 & 1 & 1 \\ 1 & 1 & 1 & 1 & -1 & -1 & -1 & -1 \\ 1 & 1 & -1 & -1 & 1 & 1 & -1 & -1 \end{bmatrix} \quad \text{are 1-resilient.}
\end{equation*}

We now wish to expand our minimal size investigation to higher degrees of resilience.

\subsection{Higher degree of resilience}

We first generalize Proposition \ref{prop:B2} to $p$-resilience.
Let us define the matrices $\bar{B}_k = [I_n \hdots I_n\ D]$ composed of $k$ identity matrices and a column vector $D = \frac{1}{\sqrt{n}}[1 \hdots 1]^\top$.

\begin{prop}\label{prop:B2p}
    The matrix $\bar{B}_{2p}$ is $p$-resilient.
\end{prop}

\begin{pf}
    We calculate $\underset{h\, \in\, \mathbb{S}}{\max}\ g(h)$ for all possible losses of $p$ actuators.
    
    First, assume the system loses control of $p$ columns belonging all to the identity matrices. Without loss of generality we assume losing one column per matrix. The index of the column lost in the $i^{th}$ identity matrix is $j_i \in [n]$. These columns form the matrix $C = \big[ e_{j_1} \hdots e_{j_p} \big]$, while $B$ is the remaining control matrix.
    Then, for $h = \big( h_1, \hdots, h_n \big) \in \mathbb{S}$, we have 
    \begin{equation*}
        C^\top h = \big( h_{j_1}, \hdots, h_{j_p}\big) \quad \text{so} \quad \|C^\top h\|^2 = \sum_{i=1}^p h_{j_i}^2,
    \end{equation*}
    \begin{equation*}
        \|B^\top h\|^2 = 2p\sum_{i = 1}^n h_i^2 -  \sum_{i = 1}^p h_{j_i}^2 + \Bigg(\sum_{i=1}^n \frac{h_i}{\sqrt{n}}\Bigg)^2 = 2p -  \sum_{i = 1}^p h_{j_i}^2 + \frac{1}{n}\Bigg(\sum_{i=1}^n h_i\Bigg)^2.
    \end{equation*}
     From \eqref{eq:function_g} we have $g(h) = \|C^\top h\| - \|B^\top h\|$. Then,
     \begin{align}\label{eq:g<0 p resilience}
         g(h) < 0 &\Longleftrightarrow  \sum_{i=1}^p h_{j_i}^2 < 2p - \sum_{i=1}^p h_{j_i}^2 + \frac{1}{n}\Bigg(\sum_{i=1}^n h_i\Bigg)^2 \nonumber \\
         &\Longleftrightarrow \sum_{i=1}^p h_{j_i}^2 < p + \frac{1}{2n} \Bigg(\sum_{i=1}^n h_i\Bigg)^2.
     \end{align}
    
    If $j_1 = \hdots = j_p$, and $h = e_{j_1}$, then \eqref{eq:g<0 p resilience} simplifies into $p < p + \frac{1}{2n}$, which is true. 
    
    If at least one of the $j_i$ is different from the others, then at least two different components of $h$ are present in the sum $\sum_{i=1}^p h_{j_i}^2$. Because $\|h\| = 1$, vector $h$ cannot have two components both equal to $1$, at least one of them is strictly inferior to $1$. Assume without loss of generality that $h_{j_1} < 1$. Because $\|h\| = 1$, we also have $h_{j_i} \leq 1$. Thus, $h_{j_1}^2 + \sum_{i=2}^p h_{j_i}^2 < 1 + \sum_{i=2}^p h_{j_i}^2 \leq 1 + (p-1) = p$.
        
    Another possible case, is that $j_1 = \hdots = j_p$ but $h \neq e_{j_1}$. Because $\|h\| = 1$, $h_{j_1} < 1$ otherwise we would have $h = e_{j_1}$. Then, $\sum_{i=1}^p h_{j_i}^2 = p h_{j_1}^2 < p$. These were the only two other possible cases and in each of them some $h_{j_i} < 1$, so the left hand side of \eqref{eq:g<0 p resilience} is strictly smaller than $p$, so the inequality also holds true. Overall $g(h) < 0$ for all $h \in \mathbb{S}$ and all choice of columns $j_1, \hdots, j_p \in [n]$.
    
    \vspace{2mm}
    The other possible loss of control is when $\bar{B}_{2p}$ loses $p-1$ columns among the identity matrices and the last column $D$. Then,
    \begin{equation*}
        g(h) = \sqrt{\sum_{i = 1}^{p-1} h_{j_i}^2 + \frac{1}{n}\Bigg(\sum_{i=1}^n h_i\Bigg)^2} - \sqrt{2p - \sum_{i = 1}^{p-1} h_{j_i}^2}\ .
    \end{equation*}
    Since $\|h\| = 1$, $h_{j_i}^2 \leq 1$ for all $i \in [p$-$1]$. We use the Cauchy-Schwarz inequality \citep{math}
    \begin{equation*}
        \left( \sum_{i=1}^n h_i \right)^2 \leq \left( \sum_{i=1}^n h_i^2 \right) \left( \sum_{i=1}^n 1^2 \right) = \|h\|^2\ n = n.
    \end{equation*}
    Then,
    \begin{equation*}
        g(h) \leq \sqrt{p-1 + \frac{1}{n} n} - \sqrt{2p - (p-1)} \leq \sqrt{p} - \sqrt{p+1} < 0.
    \end{equation*}
    Therefore, in both cases $\underset{h\, \in\, \mathbb{S}}{\max}\ g(h) < 0$. Proposition \ref{prop:B_bar_g<0} then states that $\bar{B}_{2p}$ is $p$-resilient.$\ \blacksquare$
\end{pf}

We can also extend Proposition \ref{prop:V_ortho_resilient} to 2-resilient matrices with a consequential increase in the calculations required.

\begin{prop}\label{prop:V_ortho_2resilient}
    Any matrix $V \in \mathbb{R}^{n \times m}$ where $m \geq 4n+1$    which has orthonormal rows and whose columns have all the same norm, with at least two columns being collinear, is 2-resilient.
\end{prop}
\begin{pf}
    Similarly as in the proof of Proposition \ref{prop:V_ortho_resilient} the columns have a squared norm of $\|C\|^2 = \frac{n}{m}$. 
    We extract any two columns $C_1$ and $C_2$ from $V$ to form $C$, the remaining part of $V$ is named $B$. Since $C = \big[ C_1\ C_2 \big]$, we have $CC^\top = C_1 C_1^\top + C_2 C_2^\top$.

    The singular values $\sigma^{C^\top}$ of $C^\top$ are defined as the square roots of the eigenvalues $s$ of $CC^\top$. Therefore we calculate $s = \big(\sigma^{C^\top}\big)^2$ to use Proposition \ref{prop:max sing val p-resilience}. From the matrix determinant lemma \citep{matrices},
    \begin{align*}
        0 &= \det\big(s I_n - CC^\top\big) = \det\big(s I_n - C_1 C_1^\top - C_2 C_2^\top \big) \\
        &= \big(1 - C_2^\top\big(sI_n - C_1 C_1^\top \big)^{-1} C_2 \big) \det\big(sI_n - C_1 C_1^\top\big).
    \end{align*}
    If $\det\big(sI_n - C_1 C_1^\top \big) = 0$, then the resulting eigenvalue is either $0$ or $\|C_1\|^2 = \frac{n}{m}$ by \citep{matrix_computations}.
    To investigate when the other term goes to zero, we develop the inverse into a Neumann series \citep{matrix_computations} for $s$ such that $\left\| \frac{C_1 C_1^\top}{s} \right\| < 1$: 
    \begin{align}\label{eq:Neumann_series}
       s\big(sI_n - C_1 C_1^\top \big)^{-1} &= \Bigg( I_n - \frac{C_1 C_1^\top}{s} \Bigg)^{-1} = \sum_{p=0}^\infty \Bigg( \frac{C_1 C_1^\top}{s} \Bigg)^p \\
       &= I + \sum_{p=1}^\infty \frac{1}{s^p} C_1 \big( C_1^\top C_1 \big)^{p-1} C_1^\top = I + \frac{C_1 C_1^\top}{s} \sum_{p=1}^\infty \Bigg( \frac{\|C_1\|^2}{s} \Bigg)^{p-1} \nonumber \\
       &= I + \frac{C_1 C_1^\top}{s} \frac{1}{1 - \frac{\|C_1\|^2}{s}} = I + \frac{C_1 C_1^\top}{s - \|C_1\|^2}. \nonumber
    \end{align}
    \begin{align*}
        \text{Then,} \qquad &\big(1 - C_2^\top\big(sI_n - C_1 C_1^\top \big)^{-1} C_2 \big) = 0 \quad \Longleftrightarrow \quad s =  C_2^\top s\big(sI_n - C_1 C_1^\top \big)^{-1} C_2 \\
        \Longleftrightarrow &\quad C_2^\top \left( I + \frac{C_1 C_1^\top}{s - \|C_1\|^2} \right) C_2 = s = \|C_2\|^2 + \frac{\big(C_1^\top C_2\big)^2}{s - \|C_1\|^2}  \\
        \Longleftrightarrow &\quad s^2 - \big(\|C_1\|^2 + \|C_2\|^2 \big)s + \|C_1\|^2 \|C_2\|^2 - \big(C_1^\top C_2\big)^2 = 0.
    \end{align*}
    Recall that $\|C_1\|^2 = \|C_2\|^2 = \frac{n}{m}$. Then the previous equation becomes
    \begin{equation*}
        s^2 - \frac{2n}{m} s + \frac{n^2}{m^2} - \big(C_1^\top C_2\big)^2 = 0.
    \end{equation*}
    The maximal root of this quadratic equation is
    \begin{equation}\label{eq:s_max}
            s_{max} = \frac{n}{m} + \big| C_1^\top C_2 \big|.
    \end{equation}
   This expansion is only valid for the case where $s$ satisfies $\left\|\frac{C_1 C_1^\top}{s} \right\| < 1$.
   We note that $\|C_1 C_1^\top \| = \lambda_{max}(C_1 C_1^\top) = \|C_1\|^2 = \frac{n}{m}$, from \citep{matrix_computations}. Therefore, in the other case $s \leq \frac{n}{m}$. From \eqref{eq:s_max} we deduce that $s_{max}$ is the maximal eigenvalue of $CC^\top$.
   
   \vspace{1mm}
   The matrix $C$ maximizing $s_{max}$ is the one composed of two collinear columns of $V$. Indeed, by the Cauchy-Schwarz inequality $\big| C_1^\top C_2 \big| \leq \|C_1\|\ \|C_2\|$, and the equality only happens when $C_1$ and $C_2$ are collinear. In that case, $s_{max} = \frac{2n}{m}$.
    
    \vspace{1mm}
    Then, the resilience condition of Proposition \ref{prop:max sing val p-resilience} is equivalent to $2 s_{max} < 1$, i.e., $m \geq 4n + 1$. Thus, $V$ is 2-resilient. $\quad \blacksquare$
\end{pf}

Note that two collinear columns of same norm are either the same or opposites. Proposition \ref{prop:V_ortho_2resilient} thus deals with the case where at least one actuator of the system is doubled.

With the guidelines provided by Proposition \ref{prop:V_ortho_2resilient} we produce an example of a 2-resilient matrix $V$ of size $2 \times 10$:
\begin{equation*}
    V = \begin{bmatrix} 1 & 1 & 1 & 1 & 1 & 1 & 1 & 1 & 1 & 1\\ 1 & 1 & 1 & 1 & 1 & -1 & -1 & -1 & -1 & -1 \end{bmatrix}\ .
\end{equation*}

With Proposition \ref{prop:B2p} we can generate $p$-resilient matrices of size $n \times (2pn+1)$. For $p = 1$ it corresponds to $n \times (2n+1)$, which is the minimal size for 1-resilient matrices. For $p = 2$, we obtain a matrix with $4n+1$ columns, which is consistent with the minimal size detailed in Proposition \ref{prop:V_ortho_2resilient}.

In order to determine the minimal size of a $p$-resilient matrix, with $p \geq 2$, the only missing result is an equivalent of Theorem \ref{thm:min_size} for higher degrees of resilience.

However, the process employed in the first proof of Theorem \ref{thm:min_size} does not scale well with the degree of resilience. Indeed, the fact that $\sum \det(F_i) = 0$, when $m = 2n$ cannot be generalized to $p \geq 2$. 

As for the second proof, the calculations are already significantly more complex for $p = 2$ as can be seen in the proof of Proposition \ref{prop:V_ortho_2resilient}. Without the assumption of same column norm for the case $p = 2$ the calculations do not even reach a conclusion.
For $p \geq 3$, the calculations become even more cumbersome. The Neumann series \eqref{eq:Neumann_series} becomes
\begin{equation*}
    s\left( sI_n - \sum_{j=1}^{p-1} C_j C_j^\top \right)^{-1} = \sum_{k=0}^\infty \Bigg( \sum_{j=1}^{p-1} \frac{C_j C_j^\top}{s} \Bigg)^k.
\end{equation*}
We would then need the multinomial formula to calculate each term of the series:
\begin{equation*}
    \Bigg( \sum_{j=1}^{p-1} C_j C_j^\top \Bigg)^k \hspace{-1mm} = \hspace{-3mm} \sum_{i_1+...+i_{p-1} = k} \begin{pmatrix} k \\ i_1,...,i_{p-1} \end{pmatrix} \prod_{j=1}^{p-1} \big(C_j C_j^\top\big)^{i_j}.
\end{equation*}
Proceeding to the separation of $\big(C_j C_j^\top\big)^{i_j}$ into a scalar part with the power $i_j -1$ and a matrix part like we did for $p = 2$ is still possible but brings numerous cross-terms that did not appear for $p = 2$. 
Because of the complexity of the calculations for $p \geq 2$, we were unable to obtain a simple necessary condition on the minimal size of such $p$-resilient matrices.

\begin{remark}
    If we based our intuition about the minimal size of $p$-resilient matrices on Theorem \ref{thm:min_size} and on Proposition \ref{prop:B2p}, then we might conjecture a minimal size of $n \times (2pn+1)$ for $p$-resilient matrices $\bar{B}$.
\end{remark}

Such a conjecture holds for $2$-resilient matrices with a state dimension $n = 1$. Indeed, let us consider $\bar{B} = \big[b_1\ b_2\ b_3\ b_4\big]$. Without loss of generality, assume that $b_3$ and $b_4$ have a greater absolute value than $b_1$ and $b_2$. When losing control of the last two columns we form $B = \big[b_1\ b_2\big]$ and $C = \big[b_3\ b_4\big]$. Then, $F = BB^\top - CC^\top = b_1^2 + b_2^2 - b_3^2 - b_4^2 \leq 0$. Therefore, there are no 2-resilient matrices of size $1 \times 4$. The minimal size of a 2-resilient matrix for $n = 1$ is then $1 \times 5$, since $\big[1\ 1\ 1\ 1\ 1\big]$ is 2-resilient.
    
\vspace{1mm}
However, we are able to generate 2-resilient matrices of size $n \times 4n$ for $n = 6$ and $n = 8$, and even of size $n \times (4n-2)$ for $n = 12$. Since these matrices are of consequent size, they can be found in the Appendix \ref{apx:matrices}. We will now provide the intuition that led us to these counterexamples.

We consider a matrix $V \in \mathbb{R}^{n \times m}$ with orthogonal rows whose only elements are $\pm 1$. Obviously, all columns have the same norm: $\|C\|^2 = n$, and the maximal singular value of $CC^\top$ defined in \eqref{eq:s_max} becomes $s_{max} = \big|C_1^\top C_2 \big| + n$, with the notations from the proof of Proposition \ref{prop:V_ortho_2resilient}. To build a 2-resilient matrix of minimal size, we need to minimize $s_{max}$.
Indeed, for these matrices the resilience condition of Proposition \ref{prop:max sing val p-resilience} becomes $2s_{max} < m$. For a small $s_{max}$, we should then be able to have a small number $m$ of columns.
To minimize $s_{max}$, $V$ should not have any collinear columns, because they would maximize the scalar product $\big| C_1^\top C_2 \big|$, as seen in the proof of Proposition \ref{prop:V_ortho_2resilient}.
    
There are $2^n$ different vectors composed of $n$ elements $\pm 1$. These vectors are only collinear with the vector of opposite sign. Thus, there are $2^{n-1}$ of such non-collinear vectors. To build a matrix with $4n$ columns, we then require $2^{n-1} \geq 4n$. The minimal dimension realizing that condition is $n = 6$.
We believe that it is impossible to build a 2-resilient matrix of $4n$ columns for $n \leq 5$.

We propose two ways of generating a 2-resilient matrix with $4n$ columns for $n \geq 6$. The first approach consists in producing all the non-collinear vectors and then selecting $4n$ of them to create a matrix with orthogonal rows. With this approach, we were able to produce a 2-resilient matrix of size $6 \times 24$, as can be seen in Appendix \ref{apx:matrices}.

The other approach uses the Hadamard matrices \citep{Hadamard}. They are square and orthogonal matrices composed of only $\pm 1$. By carefully selecting $n$ rows of a $4n \times 4n$ Hadamard matrix, it is possible to have $4n$ non-collinear columns.
We extracted $8$ chosen rows of a $32 \times 32$ Hadamard matrix and we built a 2-resilient matrix of minimal size $8 \times 32$ in Appendix \ref{apx:matrices}.

In order to generate a 2-resilient matrix with an even lower degree of overactuation, the maximal scalar product in \eqref{eq:s_max} must be made even smaller. We succeeded by taking $n = 12$ and selecting $n$ partial rows from a $4n \times 4n$ Hadamard matrix in order to obtain a 2-resilient matrix of size $n \times (4n-2)$ presented in the Appendix \ref{apx:matrices}.

Therefore the above conjecture is wrong. Its demise also explains why the proof of Theorem \ref{thm:min_size} cannot be extended to higher degrees of resilience.

It is now time to tackle Problem~\ref{prob:control synthesis}, the generation of a control law for resilient systems.

\section{Control synthesis}\label{sec:synthesis}

The definition of resilient reachability asks for the existence of a control law. A natural follow-up question is thus one of designing such a control law.
We want $u$ to drive the state to the target in spite of the undesirable input $w$. As noted at the beginning of the paper, if matrix $B$ was invertible, the control law $u = -B^{-1}Cw$ would cancel $w$. However, $B$ might not even be a square matrix. Instead, we design the control law using the Moore-Penrose pseudo-inverse of $B$ \citep{matrix_computations}.
An additional challenge in generating the adequate control law is to ensure that for all $w \in W$, the control $u$ stays in its set $U$. To do so, we make use of the resilient reachability conditions previously established and require the positive definiteness of $F = BB^\top - CC^\top$.

\begin{thm}\label{thm:u_driftless}
    If $F \succ 0$, then there exists $\alpha > 0$ such that 
    \begin{equation}\label{eq:control_law}
        u(t) := B^\top \big(B B^\top\big)^{-1}\Big(-Cw(t) + \alpha\big(x_{goal} - x(t)\big)\Big)
    \end{equation}
    drives the resilient system \eqref{eq:driftless_ODE} to its target ball $G$, and $u \in U$ for any $w \in W$.
\end{thm}
\begin{pf}
    We need to prove that $u$ is well-defined, stays in $U$ at all time and drives the system to the target.
    We assumed that measurements of the undesirable inputs are available in real-time and the state is completely observable, so the controller has access to $w(t)$ and $x(t)$. Since $F = BB^\top - CC^\top \succ 0$, obviously $BB^\top \succ 0$, so $BB^\top$ is invertible, and the control law is well-defined.
    
    If we plug the control law \eqref{eq:control_law} into the state equation \eqref{eq:driftless_ODE} we obtain
    \begin{align*}
        \dot{x} = BB^\top\big(B B^\top\big)^{-1}\Big(-Cw + \alpha\big(x_{goal} - x\big)\Big) + Cw = \alpha\big(x_{goal} - x\big).
    \end{align*}
    The solution is $x(t) = x_{goal} + e^{-\alpha t}d$, with $d = x(0) - x_{goal}$. Since $\alpha > 0$, the state $x$ converges globally exponentially to the target. Therefore, the control law is successful.
    
    We need to prove that for all $w \in W$, we have $u \in U$, i.e., that $\|u\|_{\mathcal{L}_2} \leq 1$. Note that $x_{goal} - x(t) = -e^{-\alpha t}d$, and define $\upsilon(t) := Cw(t) + \alpha e^{-\alpha t}d$, so that $u(t) = -B^\top \big(B B^\top\big)^{-1} \upsilon(t)$. Then,
    \begin{align*}
        \|u\|_{\mathcal{L}_2}^2 &= \int_0^T \|u(t)\|_{\mathbb{R}^m}^2 dt = \int_0^T u(t)^\top u(t)\ dt = \int_0^T \upsilon(t)^\top \big(B B^\top\big)^{-\top}B\ B^\top\big(B B^\top\big)^{-1}\upsilon(t)\ dt \\
        &= \int_0^T \upsilon(t)^\top \big(B B^\top\big)^{-1} \upsilon(t)\ dt.
    \end{align*}
    To simplify, let $P := \big(B B^\top\big)^{-1} \succ 0$, and expand $\upsilon(t)$ as
    \begin{align}\label{eq:terms_u^2}
        \upsilon(t)^\top P \upsilon(t) &= \underbrace{w(t)^\top C^\top P C w(t)}_{=\ T_1} + \underbrace{w(t)^\top C^\top P \alpha e^{-\alpha t}d}_{=\ T_2} + \underbrace{\alpha e^{-\alpha t}d^\top P C w(t)}_{=\ T_3} + \underbrace{\alpha^2 d^\top e^{-\alpha t} P e^{-\alpha t}d}_{=\ T_4} \nonumber \\
        &= T_1 + T_2 + T_3 + T_4.
    \end{align}
    
   The first term $T_1$ is the most complicated to bound. From the Woodbury formula \citep{matrix_computations}, we learn that $\big( I + C^\top F^{-1} C\big)$ is invertible, and we simplify the inverse of $BB^\top = F + CC^\top$. Since $F$ is invertible,
    \begin{equation*}
        P = \big( F + CC^\top \big)^{-1} \hspace{-2mm} = F^{-1} - F^{-1} C \big( I + C^\top F^{-1} C\big)^{-1} C^\top F^{-1} .
    \end{equation*}
    Now we define $D := C^\top F^{-1} C$. Then, $C^\top P C = D - D \big(I + D\big)^{-1}D$. \hfill\break
    By expanding $\big(I + D\big)^{-1}\big(I + D\big) = I$, we easily obtain $\big(I + D\big)^{-1}D = I - \big(I + D\big)^{-1}$, so that
    \begin{equation*}
        C^\top P C = D - D + D \big(I + D\big)^{-1}.
    \end{equation*}
    Similarly, from $\big(I + D\big)\big(I + D\big)^{-1} = I$, we finally obtain $C^\top \big(BB^\top\big)^{-1} C = I - \big(I + D\big)^{-1}$.
    
    \noindent Let $\lambda$ be an eigenvalue of $C^\top \big(BB^\top\big)^{-1} C$. Then
    \begin{align*}
        0 &= \det\big( \lambda I - C^\top \big(BB^\top\big)^{-1} C \big) = \det\big(\lambda I - I + \big(I + D\big)^{-1}\big) \\
        &= \det\big( (\lambda -1)(I+D)(I+D)^{-1} + I(I+D)^{-1}  \big) \\
        &= \det\big( (\lambda -1)(I+D) + I\big) \det(I+D)^{-1}.
    \end{align*}
    From the Woodbury formula we know that $(I+D)$ is invertible, so $\det(I+D))^{-1} \neq 0$.
    If $\lambda = 1$, then $\det(I) = 0$, which is absurd. Thus $\lambda \neq 1$, so we can divide by $(\lambda-1)$:
    \begin{equation*}
        0 = \det\Big( I + D + \frac{1}{\lambda -1}I\Big) = \det\Big(\frac{\lambda}{\lambda -1}I + D\Big) = (-1)^m \det\Big(\frac{-\lambda}{\lambda -1}I - D\Big) .
    \end{equation*}
    Since $D$ is symmetric, its eigenvalues are nonnegative, so $\frac{-\lambda}{\lambda -1} \geq 0$. Since $C^\top \big(BB^\top\big)^{-1} C$ is also symmetric, $\lambda \geq 0$. Therefore $\lambda -1 < 0$, i.e. $\lambda < 1$.
    Define $\lambda_M < 1$ as the maximal eigenvalue of $C^\top \big(BB^\top\big)^{-1} C$. Then,
    \begin{equation}\label{eq:T_1_multidim}
        \int_0^T  T_1 dt = \int_0^T w(t)^\top C^\top \big(BB^\top\big)^{-1} C w(t)\ dt \leq \int_0^T  w(t)^\top \lambda_M w(t)\ dt = \lambda_M \|w\|_{\mathcal{L}_2}^2 \leq \lambda_M.
    \end{equation} 
    We can now tackle the integral of the second term of \eqref{eq:terms_u^2}:
    \begin{equation}\label{eq:integral_T_2}
        \int_0^T T_2\ dt = \int_0^T \alpha w(t)^\top C^\top P e^{-\alpha t}d\ dt = \alpha \int_0^T w(t)^\top e^{-\alpha t}\ dt\ C^\top P d.
    \end{equation}
    Then, we calculate the norm of the integral term in \eqref{eq:integral_T_2} and use the Cauchy-Schwarz inequality to bound it:
    \begin{align*}
        \left|\left| \int_0^T w(t)^\top e^{-\alpha t}dt \right|\right|_{\mathbb{R}^m} &= \sqrt{\sum_{i=1}^m \left(  \int_0^T w_i(t) e^{-\alpha t} dt \right)^2 }  \leq \sqrt{ \sum_{i=1}^m \left(  \int_0^T \hspace{-3mm} w_i^2(t) dt\right)\left(\int_0^T \hspace{-3mm} e^{-2\alpha t} dt \right) } \\
        &\leq \sqrt{\Bigg[ \frac{e^{-2\alpha t}}{-2\alpha} \Bigg]_0^T \int_0^T \sum_{i=1}^m w_i^2(t) dt } \ =\ \sqrt{\frac{1 - e^{-2\alpha T}}{2\alpha} }\ \|w\|_{\mathcal{L}_2}.
    \end{align*}
    Thus,
    \begin{equation}\label{eq:T_2}
         \int_0^T \hspace{-2mm} T_2\ dt \leq \sqrt{\frac{\alpha}{2}} \|C^\top P d\|\ \|w\|_{\mathcal{L}_2}. %\sqrt{\big(1 - e^{-2\alpha T}\big)}
    \end{equation}
    The same process is applied to $T_3$, and results in the same upper bound:
    \begin{equation}\label{eq:T_3}
         \int_0^T \hspace{-2mm} T_3\ dt \leq \sqrt{\frac{\alpha}{2}} \|d^\top P C\|\ \|w\|_{\mathcal{L}_2}. %\sqrt{\big(1 - e^{-2\alpha T}\big)}
    \end{equation}
    We also simplify the integral of the fourth term of \eqref{eq:terms_u^2}:
    \begin{align}\label{eq:T_4}
        \int_0^T T_4 &= \int_0^T \alpha^2 d^\top e^{-\alpha t} P e^{-\alpha t}d\ dt = \alpha^2 d^\top P d \int_0^T e^{-2\alpha t}\ dt = \alpha^2 d^\top P d \Bigg[ \frac{e^{-2\alpha t}}{-2\alpha}\Bigg]_0^T \nonumber \\
        &= \frac{\alpha}{2} d^\top P d \big( 1 - e^{-2\alpha T}\big) \leq \frac{\alpha}{2}\ d^\top P d .
    \end{align}
    Then, we combine \eqref{eq:T_1_multidim}, \eqref{eq:T_2}, \eqref{eq:T_3} and \eqref{eq:T_4}:  
    \begin{equation}\label{eq:alpha}
        \|u\|_{\mathcal{L}_2}^2\ \leq\ \frac{\alpha}{2}\ d^\top P d + 2\sqrt{\frac{\alpha}{2}} \|C^\top P d\| + \lambda_M.
    \end{equation}
    Since $\lambda_M < 1$, and $d$, $P$ and $C$ are constant, we can choose $\alpha$ small enough so that the right hand side of \eqref{eq:alpha} is smaller than $1$, which finally leads to $\|u\|_{\mathcal{L}_2}^2 \leq 1$, i.e. $u \in U$. $\quad \blacksquare$
\end{pf}

The proof of Theorem~\ref{thm:u_driftless} provides a constructive method of finding $\alpha$ satisfying the claim of the theorem. The maximum $\alpha$ satisfying Theorem~\ref{thm:u_driftless} and thus ensuring the fastest convergence to $x_{goal}$ is given by 
\begin{equation}\label{eq:alpha_star}
    \alpha^* = 2\ \frac{\big(\sqrt{b^2 + (1-\lambda_M) a} - b\big)^2}{a^2}, \qquad \text{with} \quad a = d^\top P d \quad \text{and} \quad b = \|C^\top P d \|.
\end{equation}

Theorem \ref{thm:u_driftless} gives an intuitive validation of the work developed in the previous sections. Indeed, we established that resilient reachability implies $F \succ 0$.
From Theorem \ref{thm:u_driftless}, we see that such a condition is indeed sufficient to build a control law of the form \eqref{eq:control_law}.

The positive definiteness of $F$ brings two results.
The part $BB^\top \succ 0$ guarantees the existence of $u$. But $BB^\top$ is more than just positive definite, in fact $BB^\top \succ CC^\top$. This relation ensures that $u$ of the form \eqref{eq:control_law} remains within the bounds of $U$ even when $w$ is maximal.

\vspace{2mm}

We finally return to the general linear system \eqref{eq:ODE}. We will show that a control law similar to \eqref{eq:control_law} can be used if matrix $A$ is not overly unstable. The intuition is that the magnitude of $u$ in excess of $w$ can be used to counteract instability of $A$ to a certain extent. We formalize our intuition below.

For all $\eta > \max(Re(\lambda(A)))$, we can find $\beta > 0$ such that $\| e^{At} \| \leq \beta e^{\eta t}$ for all $t \geq 0$ \citep{exponential}. 
Using $P = \big( BB^\top \big)^{-1}$ and $\lambda_M = \max\big(\lambda( C^\top P C) \big)$ we define for all $\eta \in \mathbb{R}$ the set
\begin{equation}\label{eq:set_of_alpha}
    \mathcal{A}_\eta :=  \Big\{ \alpha > \eta :\ \lambda_M + \frac{\alpha}{\sqrt{\alpha -\eta}} \sqrt{2}\beta \|C^\top P\| \|x_0\| + \frac{\alpha^2}{\alpha - \eta} \frac{\beta^2}{2} \|P\| \|x_0\|^2 \leq 1 \Big\}.
\end{equation}
We showed in the proof of Theorem~\ref{thm:u_driftless} that $F \succ 0$ implies $\lambda_M < 1$. Then, taking $\alpha$ sufficiently small would satisfy the condition of \eqref{eq:set_of_alpha}, as long as $\eta$ is even smaller. There is of course a trade-off here because taking $\eta$ very close to $\max(Re(\lambda(A)))$ leads to a larger $\beta$ and thus requires an even smaller $\alpha$ to satisfy the inequality in \eqref{eq:set_of_alpha}.

\begin{thm}\label{thm:u_drift}
    If $F \succ 0$ and if there exists $\eta > \max(Re(\lambda(A)))$ such that $\mathcal{A}_\eta$ is not empty, then for all $\alpha \in \mathcal{A}_\eta$ the control law
    \begin{equation}\label{eq:control_law_A}
        u(t) := B^\top \big(BB^\top\big)^{-1} \big( -Cw(t) - \alpha x(t) \big)
    \end{equation}
    drives the resilient system \eqref{eq:ODE} to the origin, and $u \in U$ for any $w \in W$.
\end{thm}
\begin{remark}
    In contrast with the driftless case of Theorem~\ref{thm:u_driftless}, having $F \succ 0$ is not sufficient anymore for resilience. Indeed, the existence of $\alpha \in \mathcal{A}_\eta$ in Theorem~\ref{thm:u_drift} depends on the eigenvalues of matrix $A$ having sufficiently small real part. 
\end{remark}
\begin{pf}
    When plugging control law \eqref{eq:control_law_A} into \eqref{eq:ODE}, the dynamics become $\dot x(t) = Ax(t) - \alpha x(t)$. Then, $x(t) = e^{\tilde{A}t} x_0$ with $\tilde{A} := A - \alpha I$. Since $\alpha > \eta > \max(Re(\lambda(A)))$, matrix $\tilde{A}$ is Hurwitz, which guarantees the convergence of the state to the origin. Then, we need to verify whether $u \in U$ for all $w \in W$.
    
    We first bound the state transition matrix: $\| e^{\tilde{A}t} \| = \| e^{(A - \alpha I) t} \| = \| e^{At} e^{- \alpha t} \| \leq \beta e^{\eta t} e^{-\alpha t} = \beta e^{- \gamma t}$, with $\gamma := \alpha - \eta > 0$.
    Now, we can proceed as in the proof of Theorem~\ref{thm:u_driftless}. 
    Since $F \succ 0$, we have $P \succ 0$ and $\|u\|_{\mathcal{L}_2}^2 = \int_0^T \nu(t)^\top P \nu(t)\, dt$, with $\nu(t) = Cw(t) + \alpha e^{\tilde{A}t} x_0$. We develop the terms of this expression as in the proof of Theorem~\ref{thm:u_driftless}:
    \begin{equation*}
        \nu(t)^\top P \nu(t) = \underbrace{w(t)^\top C^\top P C w(t)}_{=\ T_1} + 2\underbrace{w(t)^\top C^\top P \alpha e^{\tilde{A}t} x_0}_{=\ T_2} + \underbrace{\alpha^2 x_0^\top e^{\tilde{A}^\top t} P e^{\tilde{A}t}x_0 }_{=\ T_4} = T_1 + 2T_2 + T_4.
    \end{equation*}
    Note that $T_1$ is exactly the same as in \eqref{eq:terms_u^2}, so that $\int_0^T T_1\, dt \leq \lambda_M$ according to \eqref{eq:T_1_multidim}. However, in the terms $T_2$ and $T_4$ the scalar exponential $e^{-\alpha t}$ of \eqref{eq:terms_u^2} has now been replaced by a matrix exponential $e^{\tilde{A} t}$. We use H\"older's inequality \citep{math} to split the following integral:
    \begin{align*}
        \int_0^T T_2\, dt &\leq \int_0^T \big| w(t)^\top \big(C^\top P \alpha e^{\tilde{A}t} x_0\big) \big|\, dt \leq \sqrt{\int_0^T \|w(t)\|^2 \, dt} \sqrt{\int_0^T \big\| C^\top P \alpha e^{\tilde{A}t} x_0\big\|^2 \, dt } \\
        &\leq \|w\|_{\mathcal{L}_2} \|C^\top P\| \alpha \|x_0\| \sqrt{\int_0^T \beta^2 e^{-2\gamma t}\, dt} \leq \alpha \beta \|C^\top P\| \|x_0\| \sqrt{\frac{1- e^{-2\gamma T}}{2 \gamma}} \\
        &\leq \frac{\alpha}{\sqrt{\alpha -\eta}} \frac{\beta}{\sqrt{2}} \|C^\top P\| \|x_0\|,
    \end{align*}
    where we used $\|w\|_{\mathcal{L}_2} \leq 1$ since $w \in W$.
    For the term $T_4$, we have
    \begin{align*}
        \int_0^T T_4\, dt &\leq \alpha^2 \|x_0\|^2 \|P\| \int_0^T \beta^2 e^{-2 \gamma t}\, dt = \alpha^2 \|x_0\|^2 \|P\| \beta^2 \left( \frac{1 - e^{-2\gamma T}}{2\gamma} \right) \leq \frac{\alpha^2}{\alpha - \eta} \frac{\beta^2}{2} \|P\| \|x_0\|^2
    \end{align*}
    Then,
    \begin{equation*}
        \|u\|_{\mathcal{L}_2}^2 \leq \lambda_M + \frac{\alpha}{\sqrt{\alpha -\eta}} \sqrt{2}\beta \|C^\top P\| \|x_0\| + \frac{\alpha^2}{\alpha - \eta} \frac{\beta^2}{2} \|P\| \|x_0\|^2.
    \end{equation*}
    Since we assumed that $\alpha \in \mathcal{A}_\eta$, we have $\|u\|_{\mathcal{L}_2} \leq 1$ according to \eqref{eq:set_of_alpha}. Hence, $u \in U$.    $\qquad \blacksquare$
\end{pf}

Note that the set $\mathcal{A}_\eta$ depends on $\|x_0\|$. Therefore, the further away the initial state is, the less instability can be counteracted by the control law.
From Theorem \ref{thm:u_drift} we can easily derive a sufficient condition for resilience and confirm our intuition about stable systems.

\begin{cor}\label{cor:sufficient_A}
    If $A$ is Hurwitz and $\bar{B}$ is $p$-resilient, then the system $\dot x = Ax + \bar{B}u$ is also $p$-resilient for $x_{goal} = 0$.
\end{cor}
\begin{pf}
    Because $\bar{B}$ is $p$-resilient, we can remove any $p$ columns of $\bar{B}$ and obtain $F \succ 0$. Since $A$ is Hurwitz, all its eigenvalues have a negative real part, and thus we can pick $\eta < 0$ such that $\eta > \max(Re(\lambda(A)))$. Then, $\alpha = 0 \in \mathcal{A}_\eta$ because we showed in the proof of Theorem~\ref{thm:u_driftless} that $F \succ 0$ implies $\lambda_M < 1$. Thus, we can apply Theorem~\ref{thm:u_drift} where the control law \eqref{eq:control_law_A} drives the state to $x_{goal} = 0$.  $\quad \blacksquare$
\end{pf}

We have here obtained a simple resilience condition for non-driftless systems.
We now proceed to computationally confirm the above theoretical results.

\section{Numerical example}\label{sec:examples}

To validate our theory, we consider the ADMIRE fighter jet model developed by the Swedish Defense Research Agency \citep{FOI_Admire}. The ADMIRE model has already served as an application case in several control frameworks \citep{ADMIRE_1}, \citep{ADMIRE_2}.

We explore three different scenarios featuring the fighter jet. First, we investigate the resilience of the simplified model used in \citep{ADMIRE_1}. We also use this model as a benchmark to compare our approach with a robust control method. We finally study the resilience of a more advanced driftless dynamics model of the aircraft.

\begin{figure}[htbp!]
    \centering
    \includegraphics[scale=0.5]{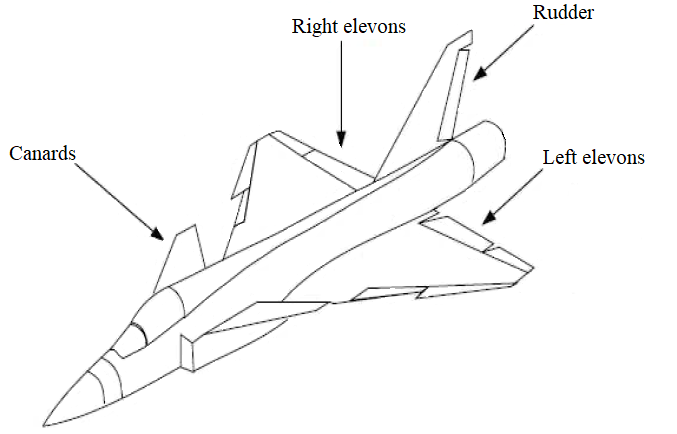}
    \caption{The ADMIRE fighter jet model. Image modified from \citep{ADMIRE_2}.}
    \label{fig:admire}
\end{figure}

\subsection{Resilience of a fighter jet}\label{subsec:Admire}

We consider only four of the actuators of the jet: the canard, the left and right elevons and the rudder, as depicted on Figure~\ref{fig:admire}. With these control surfaces, the pilot can directly affect the angular acceleration in roll, pitch and yaw.

The nominal linearized dynamics of the jet established in \citep{ADMIRE_1} are $\dot{x} = Ax + \bar{B}u$, with the state vector $x$ gathering the angular velocities in roll, pitch and yaw (rad/s): 
\begin{align*}
     x = \begin{bmatrix} p \\ q \\ r \end{bmatrix} \quad A = \begin{bmatrix} -0.997 & 0 & 0.618 \\ 0 & -0.506 & 0 \\ -0.094 & 0 & -0.213 \end{bmatrix} \quad  \bar{B} = \begin{bmatrix} 0 & -4.242 & 4.242 & 1.487 \\    1.653 & -1.274 & -1.274 & 0.002 \\ 0 & -0.281 & 0.281 & -0.882 \end{bmatrix}.
\end{align*}
Note that the system is stable since the eigenvalues of $A$ have negative real parts. 
The inputs of the system are the deflections of the control surfaces: $u_c$ for the canard wings, $u_{re}$ and $u_{le}$ for the right and left elevons, and $u_r$ for the rudder. They are mechanically constrained: 
\begin{equation}\label{eq:u}
    u = \big( u_c,\ u_{re},\ u_{le},\ u_r \big) \quad \text{with} \quad u_c \in [-25, 55] \frac{\pi}{180}, \quad \text{and} \quad u_{re}, u_{le}, u_r \in [-30, 30] \frac{\pi}{180}.
\end{equation}

Consider the scenario in which, after sustaining damage (e.g., during air combat), one of the control surfaces of the fighter jet stops responding to the commands. This surface is now producing undesirable inputs. The pilot wants to minimize the aircraft roll, pitch and yaw rates, so the target is a ball of radius $0.1$ centered around the origin, $x = 0$.

By looking at the matrix $\bar{B}$ we can build our intuition on the resilience of the system. The first column represents the effect of the canard and only modifies the pitch rate of the aircraft. This actuator can be counteracted by the combined actions of both elevons, because $1.2735 + 1.2735 > 1.6532$. The elevons can counteract each other in terms of roll but doing so would induce a high pitching moment that cannot be counteracted. The yawing moment produced by the rudder cannot be counteracted by the other actuators: $0.8823 > 0.2805 + 0.2805$. Therefore, our intuition states that the fighter jet is only resilient to the loss of control authority over the canard.

We check whether the matrix $F = BB^\top - CC^\top \succ 0$ for each of the four possible actuator losses. Table \ref{tab:eigenvalues F} gathers the minimal eigenvalues of $F$ for the four cases. As predicted by our intuition, the jet is only resilient to the loss of control authority over the canard.

\begin{table}[!htbp]
\renewcommand{\arraystretch}{1.3}
\caption{Minimal eigenvalue of $F$ for each actuator losses}
\label{tab:eigenvalues F}
\centering
\begin{tabular}{|c||c|c|c|c|}
\hline
Loss of control of: & Canards & Right elevon & Left elevon & Rudder \\
\hline
$\min \lambda(F)$ & 0.51 & -8.5 & -8.5 & -1.0 \\
\hline
\end{tabular}
\end{table}

We study more in-depth the loss of control over the canard with Theorem \ref{thm:u_drift}. We reuse the notations employed in the proof and after some calculations we obtain: $\lambda_M = 0.8426 < 1$, $\max(Re(\lambda(A))) = -0.259 < \alpha^*$, so the control law \eqref{eq:control_law_A} should work. 

We simulate our system on MATLAB with $ode45$ on the time interval $[0, 25]$. We generate $w$ as a stochastic signal between the bounds of $u_c$ defined in \eqref{eq:u}, i.e., $w(t) \in [-25, 55]\frac{\pi}{180}$ for $t \in [0, 25]$. If $\|w\|_{\mathcal{L}_2} > 1$, we divide $w$ by its $\mathcal{L}_2$ norm so that once normalized, $\|w\|_{\mathcal{L}_2} = 1$. If instead we initially had $\|w\|_{\mathcal{L}_2} \leq 1$, then we keep $w$ as is. In order to respect the constraints \eqref{eq:u} we add a saturation to the control law \eqref{eq:control_law_A} and to the LQR feedback control law $u_{LQR} = -Kx$ that we use as a reference. On MATLAB we obtain
\begin{equation*}
    K = \begin{bmatrix} -0.5825 & -0.5358 & -0.1659 \\ 0.5826 & -0.5360 & 0.1653 \\ 0.2198 & 0.0007 & -0.7564 \end{bmatrix}, \quad \text{with} \quad Q = \begin{bmatrix} 1 & 0 & 0 \\ 0 & 1 & 0 \\ 0 & 0 & 1 \end{bmatrix}, \quad \text{and} \quad R = 1.
\end{equation*} 

\begin{figure}[htbp!]
    \centering
    \begin{subfigure}{0.49\textwidth}
        \centering
        \includegraphics[scale=0.5]{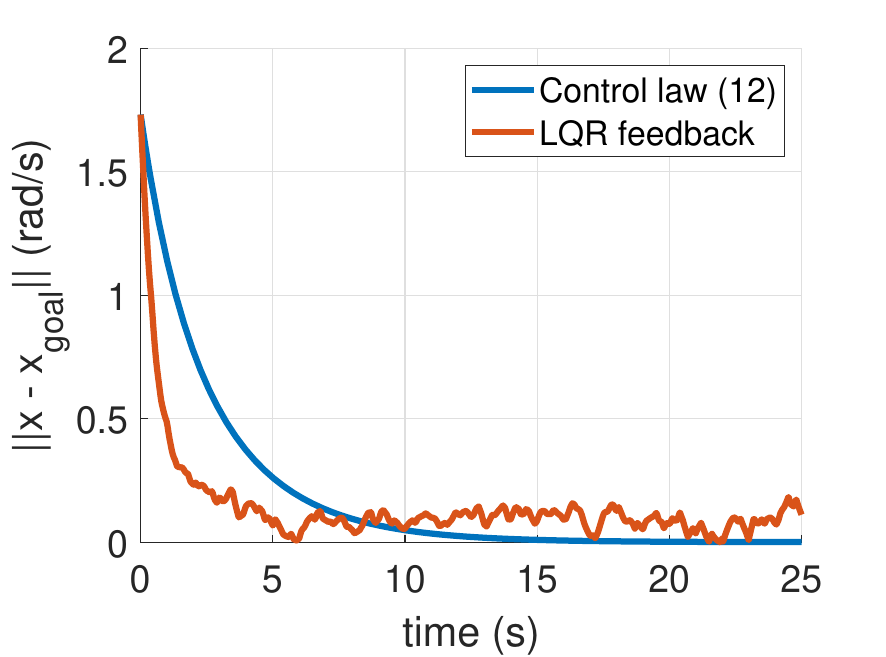}
       \caption{Distance of the state from the origin.}
        \label{fig:exp_cv}
    \end{subfigure}
    \begin{subfigure}{0.49\textwidth}
        \centering
        \includegraphics[scale=0.5]{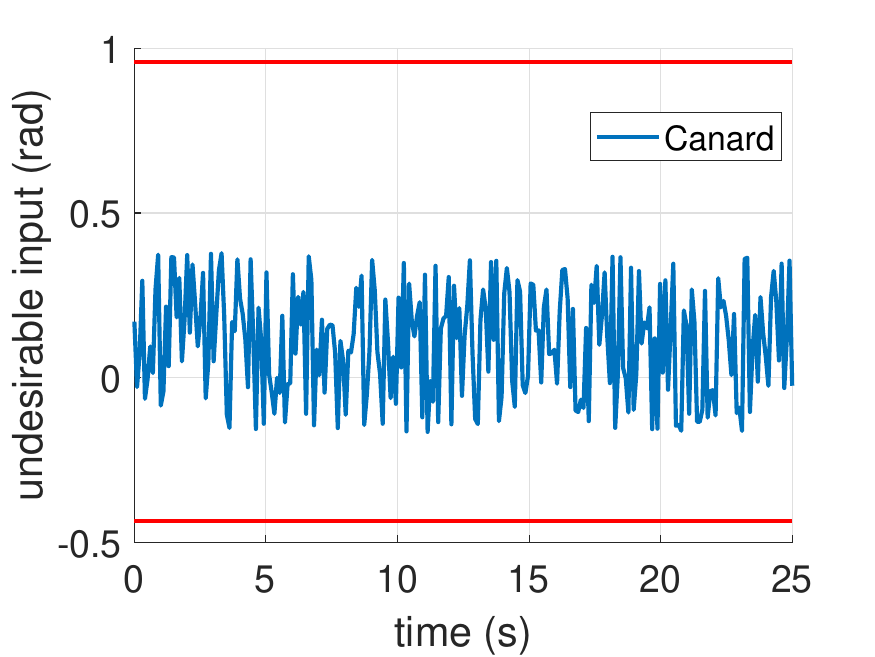}
        \caption{Undesirable canard inputs.}
        \label{fig:canard}
    \end{subfigure}
    \caption{State evolution with the two controllers for undesirable canard inputs.}
    \label{fig:2}
\end{figure}

As predicted, the state converges exponentially from $x_0 = (1,\ 1,\ 1)\ rad/s$ to the origin, as shown by the blue curve in Figure~\ref{fig:2}(\subref{fig:exp_cv}). With the LQR feedback unaware of the undesirable input, the state does not converge to the origin, as shown in red in Figure~\ref{fig:2}(\subref{fig:exp_cv}).
As can be seen on Figure~\ref{fig:2}(\subref{fig:canard}), the undesirable input has a high variation and an amplitude non-negligible compared to the controlled inputs. It is not reaching its upper and lower bound because of the $\mathcal{L}_2$ normalization we operated.

\begin{figure}[htbp!]
    \centering
    \begin{subfigure}{0.49\textwidth}
        \centering
        \includegraphics[scale=0.5]{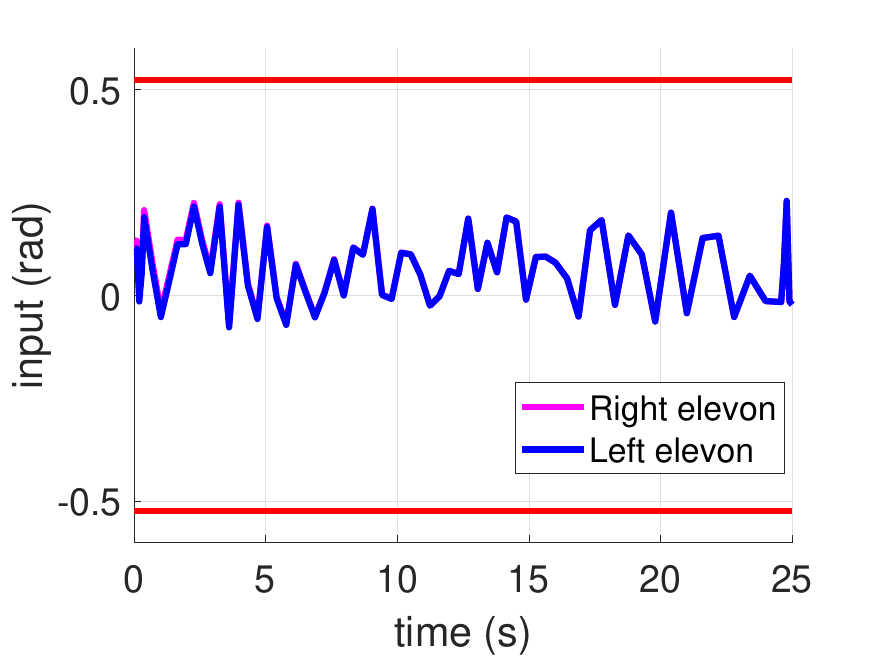}
        \caption{Control law \eqref{eq:control_law}.}
        \label{fig:elevons}
    \end{subfigure}
    \begin{subfigure}{0.49\textwidth}
        \centering
        \includegraphics[scale=0.5]{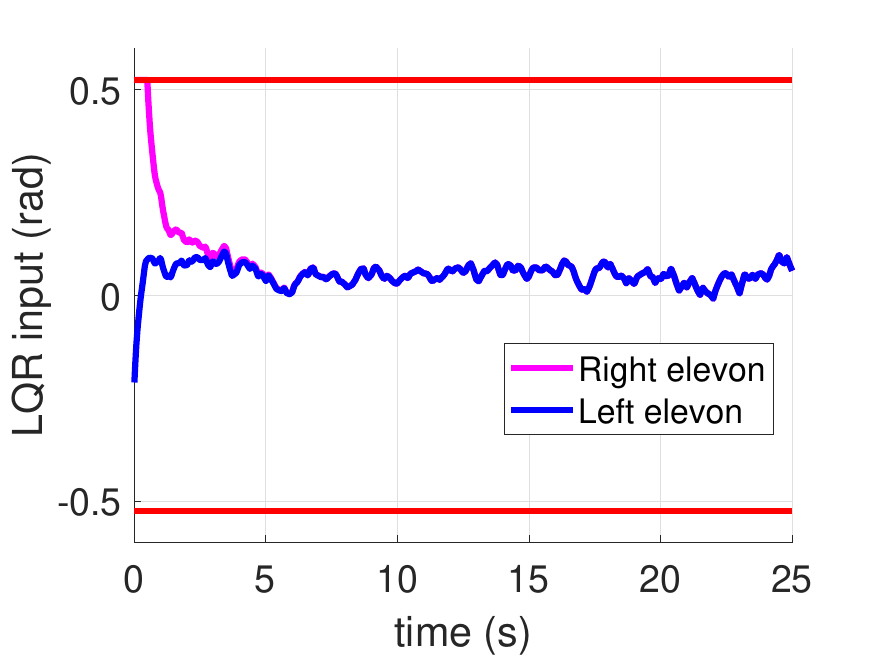}
        \caption{LQR feedback control.}
        \label{fig:LQR_elevons}
    \end{subfigure}
    \caption{Right and left elevons inputs for the two controllers.}
    \label{fig:3}
\end{figure}

\begin{figure}[htbp!]
    \centering
    \begin{subfigure}{0.49\textwidth}
        \centering
        \includegraphics[scale=0.54]{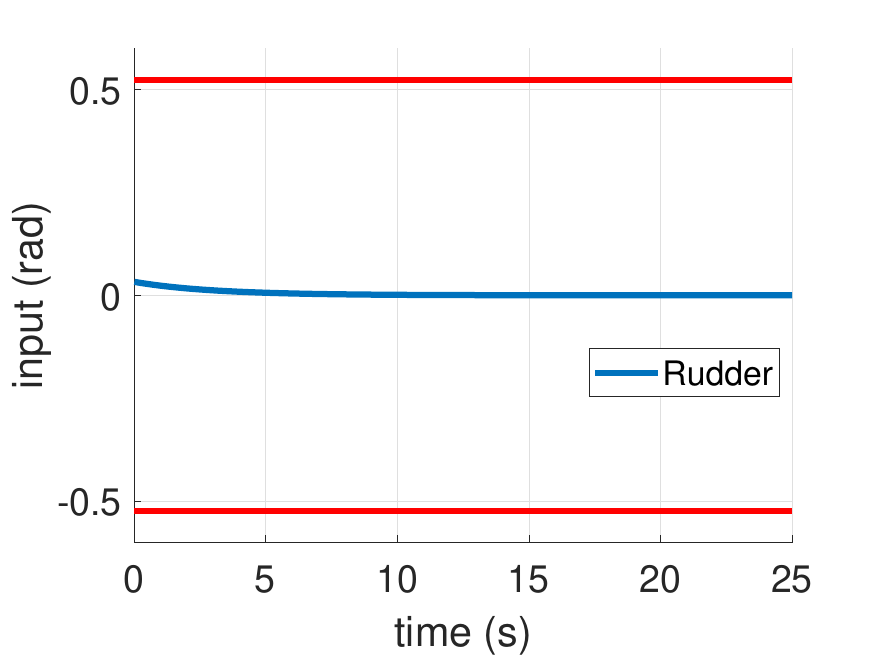}
        \caption{Control law \eqref{eq:control_law}.}
        \label{fig:rudder}
    \end{subfigure}
    \begin{subfigure}{0.49\textwidth}
        \centering
        \includegraphics[scale=0.54]{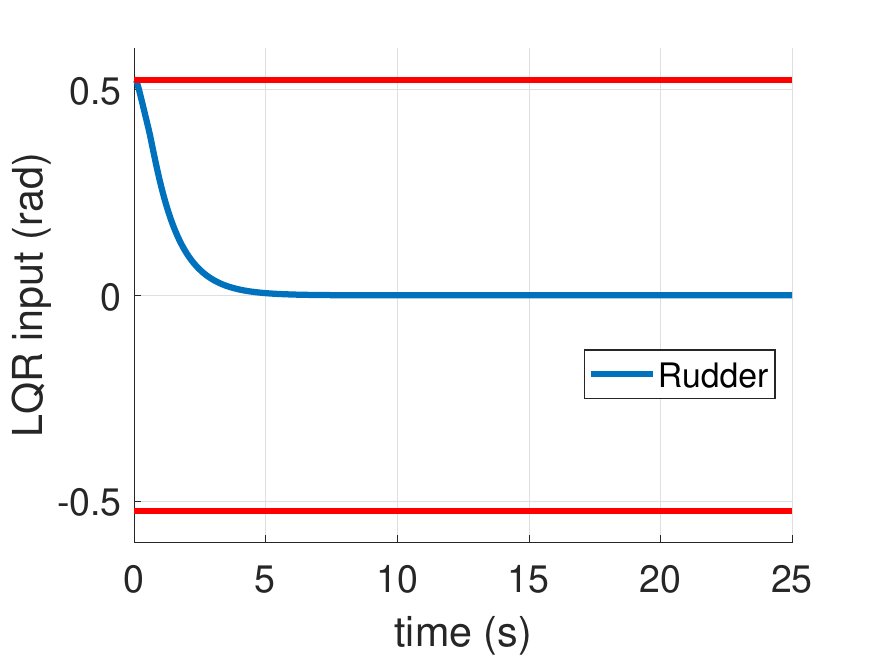}
        \caption{LQR feedback control.}
        \label{fig:LQR_rudder}
    \end{subfigure}
    \caption{Rudder inputs for the two controllers.}
    \label{fig:4}
\end{figure}

The control strategies employed by our two controllers are very different, as illustrated by the differences between Figure~\ref{fig:3}(\subref{fig:elevons}) and \ref{fig:3}(\subref{fig:LQR_elevons}), and between Figure~\ref{fig:4}(\subref{fig:rudder}) and \ref{fig:4}(\subref{fig:LQR_rudder}).
The LQR input is initially saturated as can be seen on Figures~\ref{fig:3}(\subref{fig:LQR_elevons}) and \ref{fig:4}(\subref{fig:LQR_rudder}).

If the pilot loses control authority over any one of the elevons, then $F$ is not positive definite, but $BB^\top$ is invertible.
The control law \eqref{eq:control_law_A} is still well-defined, so it can be implemented, but for some $w \in W$ the control is not admissible: $u_w \notin U$. 
% We cannot guarantee that it will overcome the undesirable inputs.

If the pilot loses control of the rudder, $BB^\top$ is not invertible, so the control law \eqref{eq:control_law_A} is not well-defined. The jet cannot be guaranteed to be able to reach the desired target.

\subsection{Comparison with robust control}\label{subsec:robust}

To illustrate the strength of our approach in the considered scenario, we compare our results with those of classical robust control.

Let us first recall the differences in assumptions between robust control and resilient reachability.
A control law is said to be robust if it drives the state to the target \textit{whatever the disturbance is}, i.e., there exists a control law $u$ such that for all undesirable input $w$, we have $x(T) \in G$.
On the other hand, resilient reachability considers a controller \textit{aware of the undesirable input}, i.e., for all $w$, there exists a control law $u_w$ such that $x(T) \in G$.

In our setting, the undesirable input is produced by an actuator belonging to the system. With sensors measuring the output of each actuators and a fault-detection mechanism, it is reasonable to assume that $w$ can be measured.
Then, the resilient controller has access to more information than a robust controller, and should perform better.

We choose the robust control approach developed in \citep{kurzhanski2002}. Its objective is to approximate the closed-loop reach set $\mathcal{X}[T]$ with internal and external ellipsoids. The reach set gathers the states $x_{goal} \in \mathbb{R}^n$ for each of which there exists a control law such that, whatever the undesirable input is, $x(T) \in \mathcal{B}(x_{goal}, \mu)$ for a certain radius $\mu \geq 0$. 

We compare the precision of our approach with \citep{kurzhanski2002} based on the size of the smallest target ball guaranteed to be reached.
The application case is the ADMIRE model with drift studied in the previous subsection \ref{subsec:Admire}. We assume that the pilot loses control authority over the canards.

% The undesirable input $w$ follows a uniform distribution within the defective's actuator range.
The resilient inputs have $\mathcal{L}_2$ bounds. However, the robust control inputs $u$ must be bounded by an ellipsoid. To make the comparison as fair as possible, we choose the maximal ellipsoid within the actuators range \eqref{eq:u}.
For the details of its construction we refer the reader to Appendix \ref{apx:robust}.

We now need to calculate the radius $\mu$ of the smallest robustly reachable target. We compute only the tight ellipsoidal internal approximation of the closed-loop reach set: \break
$\mathcal{E}\big(x_-(T), X_-(T)\big)\ \subseteq\ \mathcal{X}[T]$. 
% Indeed, if a state belongs to the internal ellipsoid it is guaranteed to be robustly reachable, because included in the reach set.
% If $\mu$ is chosen too small compared to the magnitude of the undesirable inputs, the robust control law cannot guarantee to drive the state to the target ball.
% In that case, the shape matrix $X_-(T)$ of the ellipsoid is not positive definite. Therefore, we look at the minimum target radius $\mu$ for which $X_-(T) \succ 0$, and we obtain $\mu = 5.9$.
We numerically obtained $\mu = 5.9$.
Thus, the robust control law (with standard ellipsoidal approximations of the reachable set) can only guarantee to reach a target state within a radius of $5.9$. 
The initial state $x_0 = (1,\ 1,\ 1)$ was already inside that ball. Thus, the robust control cannot even guarantee that the state will get closer to the target than its initial state.

On the other hand, we know that the jet is resilient to the loss of control over the canards. Therefore, a target ball of any size is resiliently reachable. 
By having access to the undesirable input, a controller ensuring resilient reachability is then more effective than a robust controller.

\subsection{A driftless model}

The aircraft model used as previous example is very convenient for our study because of the linearization and the overactuation. However, to render the dynamics driftless, we needed a more in-depth analysis of the model. We obtained the original simulation code of the ADMIRE model from \citep{durham2017aircraft}.

For our purposes, we removed the states representing the sensor dynamics and those not directly affected by the controls from the initial 28-states model \citep{FOI_Admire}.
We also removed four of the sixteen inputs as they are negligible compared to the other inputs. 

The simulation generates a pair of matrices $A$ and $\bar{B}$ following the nominal dynamics \eqref{eq:initial_ODE}.
The effect of the matrix $A$ is negligible compared to $\bar{B}$, when considering the states $x = (V_t,\ q,\ r)$, i.e., the jet speed, pitch and yaw rates. Thus, we approximate their dynamics by a driftless system, setting $A = 0$.
% Consider state $x_i$, its dynamics follows $\dot{x}_i = row(A)_i x + row(\bar{B})_i u$. We define the ratio $r_i = \|row(\bar{B})_i\| / \|row(A)_i\|$ and obtained values for each state in different cases. The ratios depend on the Mach and altitude, but we were able to split the states in two categories. For the angle of attack, the sideslip angle and the roll rate, the ratios were all below 5 and some below 1.5, which means that the drift dynamics account for at least $20\%$ and up to $70\%$. Therefore, these states cannot be approximated with a driftless dynamics. On the other hand, the speed, pitch and yaw rate have all ratios higher than 10. 

Since the jet has a single engine, it is not resilient to its loss. For our study, we assume a guaranteed authority over the thrust command, except for the afterburners.
In the model the thrust command actuator also encompasses the afterburners. Since they account for only $20\%$ of the thrust, the corresponding column in $\bar{B}$ is scaled by $20\%$.

At Mach 0.75 and altitude 3000 m, the control matrix is
\begin{equation*}
    \bar{B}^\top = \left[\arraycolsep=1mm \begin{array}{ccc}
        -2.7 & 7.1 & -1.9 \\
        -2.7 & 7.1 & 1.9 \\
        -1.0 & -7.7 & -1.1 \\
        -1.8 & -13 & -3.0 \\
        -1.8 & -13 & 3.0 \\
        -1.0 & -7.7 & 1.1 \\
        -1.9 & 0.0 & -11 \\
        -0.8 & -0.5 & 0 \\
        -4.3 & -0.7 & 0 \\
        1.2 & 0 & 0 \\
       -71 & 1.2 & -710 \\
       -113 & -882 & 0 \end{array}\right] \hspace{1mm}
       \begin{array}{l} \text{right canard,} \\
       \text{left canard,} \\
       \text{right outboard elevon,} \\
       \text{right inboard elevon,} \\
       \text{left inboard elevon,} \\
       \text{left outboard elevon,} \\
       \text{rudder,} \\
       \text{leading edge flaps,} \\
       \text{landing gear,}\\
       \text{afterburner,} \\
       \text{yaw thrust vectoring,} \\
       \text{pitch thrust vectoring.} \end{array}
\end{equation*}
Each row of $\bar{B}^\top$ represents the effect of the actuator written on the right. All the values of the inputs are in radians except for the landing gear and the afterburner which are between $0$ and $1$.
This control matrix is not 1-resilient, because the thrust vectoring inputs are several orders of magnitude greater than any of the other inputs. For the same reason, the system is resilient to the loss of any one of the other ten actuators.

Simply removing thrust vectoring capabilities does not render the system $1$-resilient; the control of the yaw rate would then primarily depend on the rudder, hence rendering the aircraft not resilient to the loss of the rudder.

Instead of removing the thrust vectoring actuators, if their range of motion is restricted to $1.4\%$ of their current range, then $\bar{B}$ becomes resilient.
Indeed, the thrust vectoring actuators can now be counteracted by the rudder and the elevons. 
Since we reduced the magnitude of two columns of $\bar{B}$, we also had to verify that the driftless hypothesis was still valid by comparing the effects of $A$ and $\bar{B}$.
% We recompute the ratios $\|row(\bar{B})_i\| / \|row(A)_i\|$, and they are still greater than ten.

% If we consider another design point, for instance Mach 0.25 at an altitude of a 1000 m, 
% \begin{equation*}
%     \bar{B}^\top = \left[\arraycolsep=1mm \begin{array}{ccc}
%      -1.2 & 1.3 & -0.4 \\
%      -1.2 & 1.3 & 0.4 \\
%      -0.9 & -0.9 & -0.2 \\
%      -1.3 & -1.5 & -0.4 \\
%      -1.3 & -1.5 & 0.4 \\
%      -0.9 & -0.9 & 0.2 \\
%      -0.2 & 0.0 & -1.5 \\
%       0.5 & -0.1 & 0 \\
%      -0.5 & -0.1 & 0 \\
%       1.3 & 0 & 0 \\
%      -14 & 0.2 & -142 \\
%      -55 & -177 & 0 \end{array}\right]  \hspace{-1mm}
%       \arraycolsep=1mm\begin{array}{cl} (1) & \text{right canard} \\
%       (2) & \text{left canard} \\
%       (3) & \text{right outboard elevon} \\
%       (4) & \text{right inboard elevon} \\
%       (5) & \text{left inboard elevon} \\
%       (6) & \text{left outboard elevon} \\
%       (7) & \text{rudder} \\
%       (8) & \text{leading edge flaps} \\
%       (9) & \text{landing gear} \\
%       (10) & \text{afterburner} \\
%       (11) & \text{yaw thrust vectoring} \\
%       (12) & \text{pitch thrust vectoring} \end{array}
% \end{equation*}
% Even at the lowest speed in the design space, the thrust vectoring is still two orders of magnitude greater than any of the other actuator. Remark that the dividing factor required here is different from the previous case, it is around a hundred.  

We showed how to make the fighter jet resilient in terms of speed, pitch and yaw rates, by scaling down thrust vectoring and having a guaranteed thrust.
The resilience improvement by reducing the thrust vectoring might seem counterintuitive. Yet, it is explained by the fact that these actuators were too powerful to be balanced if they became uncontrolled.
While the new system is resilient, its capabilities have been reduced. For instance, reaching a target (while undamaged) would take significantly more time for the new resilient system than for the old one.

The resilience analysis developed for this fighter jet is affected by several limitations of the current state of our theory. The first and obvious limitation comes from the driftless hypothesis but is justified here by the difference of magnitude between the drift and controlled dynamics.
The most limiting hypothesis is that the controls are bounded by a $\mathcal{L}_2$ norm. Indeed, each actuator is independent of the others so a joint bound may not be appropriate. The structure of $U$ from \eqref{eq:def of U, W, G} also assumes that each actuator has a symmetric range of functioning, which makes sense for the rudder, for instance, but not for the landing gear which can only be stored or deployed. These two main limitations lead our future work directions.

\section{Conclusions and Future Work}

This paper introduced the notion of resilient systems that can withstand the loss of control over any single or multiple actuators and still guarantee to drive the state to its target. 
We established necessary and sufficient conditions to verify the resilience of a system.
We determined the minimal number of actuators required for $1$- and $2$-resilient systems. 
Further developing the theory, we established several methods to design resilient systems of any dimension and of any degree of resilience.
We then focused on control law synthesis for driftless and non-driftless systems. We proceeded to illustrate our results on a model of a fighter jet.

There are four promising avenues of future work.
Most of our work so far has concerned driftless systems. We aim to extend the theory to a broader class of dynamics.
Another direction of work concerns the type of bounds on the inputs. In this work we considered a bound on the total actuation effort of all the actuators over time. Instead, we want each actuator to have its own bound enforced at every instant.
Another useful future step is to establish a metric quantifying the resilience of a given system, for example, comparing the time required to reach a target with and without loss of control over actuators. 
Our fourth direction of future work is to investigate more complex control specifications, e.g., reach-avoid, where the system seeks to avoid parts of the state space while reaching a target.

\appendix
\appendixpage
\section{Examples of 2-resilient matrices}\label{apx:matrices}

The following matrix $\bar{B}$ of size $6 \times 24$ is 2-resilient:
\begin{equation*}
   \hspace{-2mm} \bar{B} = \left[\arraycolsep=0.8mm\def\arraystretch{0.9}\begin{array}{cccccccccccccccccccccccc}
     1 & 1 & 1 & 1 & 1 & -1 & -1 & -1 & -1 & -1 & 1 & 1 & 1 & 1 & 1 & 1 & 1 & 1 & -1 & -1 & -1 & 1 & -1 & 1 \\
     1 & -1 & 1 & 1 & 1 & -1 & 1 & 1 & 1 & 1 & -1 & -1 & 1 & 1 & 1 & 1 & 1 & -1 & 1 & -1 & 1 & -1 & 1 & 1 \\
     1 & 1 & 1 & 1 & 1 & 1 & -1 & 1 & 1 & 1 & -1 & 1 & -1 & -1 & -1 & 1 & -1 & 1 & -1 & 1 & -1 & -1 & 1 & 1 \\
     1 & 1 & -1 & 1 & 1 & 1 & 1 & -1 & 1 & 1 & 1 & 1 & -1 & 1 & 1 & -1 & -1 & -1 & 1 & -1 & -1 & 1 & 1 & -1 \\
     1 & 1 & 1 & -1 & 1 & 1 & 1 & 1 & -1 & 1 & 1 & -1 & 1 & -1 & 1 & 1 & -1 & -1 & -1 & 1 & 1 & 1 & -1 & -1 \\
     1 & 1 & 1 & 1 & -1 & 1 & 1 & 1 & 1 & -1 & 1 & 1 & 1 & 1 & -1 & -1 & 1 & 1 & 1 & 1 &  1 & -1 & -1 & -1
    \end{array}\right].
\end{equation*}

% \begin{equation*}
%      \bar{B}^\top = \left[\def\arraystretch{0.9}\begin{array}{cccccc}
%      1 & 1 & 1 & 1 & 1 & 1 \\
%      1 &-1 & 1 & 1 & 1 & 1 \\
%      1 & 1 & 1 &-1 & 1 & 1 \\
%      1 & 1 & 1 & 1 &-1 & 1 \\
%      1 & 1 & 1 & 1 & 1 &-1 \\
%      -1 &-1 & 1 & 1 & 1 & 1 \\
%      -1 & 1 &-1 & 1 & 1 & 1 \\
%      -1 & 1 & 1 &-1 & 1 & 1 \\
%      -1 & 1 & 1 & 1 &-1 & 1 \\
%      -1 & 1 & 1 & 1 & 1 &-1 \\
%       1 &-1 &-1 & 1 & 1 & 1 \\
%       1 &-1 & 1 & 1 &-1 & 1 \\
%       1 & 1 &-1 &-1 & 1 & 1 \\
%       1 & 1 &-1 & 1 &-1 & 1 \\
%       1 & 1 & -1 &  1 &  1 & -1 \\
%       1 &  1 &  1 & -1 &  1 & -1 \\
%       1 &  1 & -1 & -1 & -1 &  1 \\
%       1 & -1 &  1 & -1 & -1 &  1 \\
%       -1 &  1 & -1 &  1 & -1 &  1 \\
%       -1 & -1 &  1 & -1 &  1 &  1 \\
%       -1 &  1 & -1 & -1 &  1 &  1 \\
%       1 & -1 & -1 &  1 &  1 & -1 \\
%       -1 &  1 &  1 &  1 & -1 & -1 \\
%       1 &  1 &  1 & -1 & -1 & -1 
%     \end{array}\right].
% \end{equation*}

\vspace{-5mm}
% The following matrix $\bar{B}$ of size $$ is 2-resilient:
\begin{equation*}
    \text{The matrix $\bar{B}$ of size}\ 8 \times 32\ \text{is 2-resilient:} \quad \bar{B}^\top = \left[\def\arraystretch{0.9}\begin{array}{cccccccc}
      1 &  1 &  1 &  1 &  1 &  1 &  1 &  1 \\
      1 &  1 &  1 &  1 &  1 &  1 &  1 & -1 \\
      1 &  1 &  1 &  1 &  1 &  1 & -1 &  1 \\
      1 &  1 &  1 &  1 &  1 &  1 & -1 & -1 \\
      1 &  1 &  1 &  1 & -1 & -1 &  1 &  1 \\
      1 &  1 & -1 & -1 &  1 &  1 &  1 &  1 \\
     -1 & -1 &  1 &  1 &  1 &  1 &  1 &  1 \\
      1 &  1 &  1 &  1 & -1 & -1 &  1 & -1 \\
      1 &  1 & -1 & -1 &  1 &  1 &  1 & -1 \\
     -1 & -1 &  1 &  1 &  1 &  1 &  1 & -1 \\
      1 &  1 &  1 &  1 & -1 & -1 & -1 &  1 \\
      1 &  1 & -1 & -1 &  1 &  1 & -1 &  1 \\
     -1 & -1 &  1 &  1 &  1 &  1 & -1 &  1 \\
      1 & -1 &  1 & -1 &  1 & -1 &  1 &  1 \\
      1 & -1 & -1 &  1 &  1 & -1 &  1 &  1 \\
     -1 &  1 &  1 & -1 &  1 & -1 &  1 &  1 \\
     -1 &  1 & -1 &  1 &  1 & -1 &  1 &  1 \\
      1 & -1 & -1 &  1 & -1 &  1 &  1 &  1 \\
      1 & -1 &  1 & -1 & -1 &  1 &  1 &  1 \\
     -1 &  1 & -1 &  1 & -1 &  1 &  1 &  1 \\
     -1 &  1 &  1 & -1 & -1 &  1 &  1 &  1 \\
      1 &  1 &  1 &  1 & -1 & -1 & -1 & -1 \\
      1 &  1 & -1 & -1 &  1 &  1 & -1 & -1 \\
      1 & -1 & -1 &  1 &  1 & -1 &  1 & -1 \\
      1 & -1 &  1 & -1 &  1 & -1 &  1 & -1 \\
      1 & -1 & -1 &  1 & -1 &  1 &  1 & -1 \\
      1 & -1 &  1 & -1 & -1 &  1 &  1 & -1 \\
      1 & -1 & -1 &  1 & -1 &  1 & -1 &  1 \\
      1 & -1 & -1 &  1 &  1 & -1 & -1 &  1 \\
      1 & -1 &  1 & -1 &  1 & -1 & -1 &  1 \\
      1 & -1 &  1 & -1 & -1 &  1 & -1 &  1 \\
      1 &  1 & -1 & -1 & -1 & -1 &  1 &  1
    \end{array}\right] .
\end{equation*}

The following matrix $\bar{B}$ of size $12 \times 46$ is 2-resilient:
\begin{equation*}
    \bar{B}^\top = \left[\arraycolsep=0.9mm\def\arraystretch{0.9}\begin{array}{cccccccccccc}
      1 &  1 &  1 &  1 &  1 &  1 &  1 &  1 &  1 &  1 &  1 &  1 \\
     -1 &  1 &  1 & -1 &  1 &  1 &  1 & -1 & -1 &  1 & -1 &  1 \\
     -1 & -1 &  1 &  1 & -1 &  1 &  1 &  1 & -1 & -1 & -1 & -1 \\
     -1 &  1 & -1 &  1 &  1 & -1 &  1 &  1 & -1 &  1 & -1 &  1 \\
     -1 & -1 &  1 & -1 &  1 &  1 & -1 &  1 & -1 & -1 & -1 & -1 \\
     -1 & -1 & -1 &  1 & -1 &  1 &  1 & -1 & -1 & -1 & -1 & -1 \\
     -1 & -1 & -1 & -1 &  1 & -1 &  1 &  1 & -1 & -1 & -1 & -1 \\
     -1 &  1 & -1 & -1 & -1 &  1 & -1 &  1 & -1 &  1 & -1 &  1 \\
     -1 &  1 &  1 & -1 & -1 & -1 &  1 & -1 & -1 &  1 & -1 &  1 \\
     -1 &  1 &  1 &  1 & -1 & -1 & -1 &  1 & -1 &  1 & -1 &  1 \\
     -1 & -1 &  1 &  1 &  1 & -1 & -1 & -1 & -1 & -1 & -1 & -1 \\
     -1 &  1 & -1 &  1 &  1 &  1 & -1 & -1 & -1 &  1 & -1 &  1 \\
      1 &  1 &  1 &  1 &  1 &  1 &  1 &  1 & -1 & -1 &  1 &  1 \\
     -1 &  1 &  1 & -1 &  1 &  1 &  1 & -1 &  1 & -1 & -1 &  1 \\
     -1 & -1 &  1 &  1 & -1 &  1 &  1 &  1 &  1 &  1 & -1 & -1 \\
     -1 &  1 & -1 &  1 &  1 & -1 &  1 &  1 &  1 & -1 & -1 &  1 \\
     -1 & -1 &  1 & -1 &  1 &  1 & -1 &  1 &  1 &  1 & -1 & -1 \\
     -1 & -1 & -1 &  1 & -1 &  1 &  1 & -1 &  1 &  1 & -1 & -1 \\
     -1 & -1 & -1 & -1 &  1 & -1 &  1 &  1 &  1 &  1 & -1 & -1 \\
     -1 &  1 & -1 & -1 & -1 &  1 & -1 &  1 &  1 & -1 & -1 &  1 \\
     -1 &  1 &  1 & -1 & -1 & -1 &  1 & -1 &  1 & -1 & -1 &  1 \\
     -1 &  1 &  1 &  1 & -1 & -1 & -1 &  1 &  1 & -1 & -1 &  1 \\
     -1 & -1 &  1 &  1 &  1 & -1 & -1 & -1 &  1 &  1 & -1 & -1 \\
     -1 &  1 & -1 &  1 &  1 &  1 & -1 & -1 &  1 & -1 & -1 &  1 \\
      1 &  1 &  1 &  1 &  1 &  1 &  1 &  1 &  1 &  1 & -1 & -1 \\
     -1 &  1 &  1 & -1 &  1 &  1 &  1 & -1 & -1 &  1 &  1 & -1 \\
     -1 & -1 &  1 &  1 & -1 &  1 &  1 &  1 & -1 & -1 &  1 &  1 \\
     -1 &  1 & -1 &  1 &  1 & -1 &  1 &  1 & -1 &  1 &  1 & -1 \\
     -1 & -1 &  1 & -1 &  1 &  1 & -1 &  1 & -1 & -1 &  1 &  1 \\
     -1 & -1 & -1 &  1 & -1 &  1 &  1 & -1 & -1 & -1 &  1 &  1 \\
     -1 & -1 & -1 & -1 &  1 & -1 &  1 &  1 & -1 & -1 &  1 &  1 \\
     -1 &  1 & -1 & -1 & -1 &  1 & -1 &  1 & -1 &  1 &  1 & -1 \\
     -1 &  1 &  1 & -1 & -1 & -1 &  1 & -1 & -1 &  1 &  1 & -1 \\
     -1 &  1 &  1 &  1 & -1 & -1 & -1 &  1 & -1 &  1 &  1 & -1 \\
     -1 & -1 &  1 &  1 &  1 & -1 & -1 & -1 & -1 & -1 &  1 &  1 \\
     -1 &  1 & -1 &  1 &  1 &  1 & -1 & -1 & -1 &  1 &  1 & -1 \\
      1 &  1 &  1 &  1 &  1 &  1 &  1 &  1 & -1 & -1 & -1 & -1 \\
     -1 &  1 &  1 & -1 &  1 &  1 &  1 & -1 &  1 & -1 &  1 & -1 \\
     -1 & -1 &  1 &  1 & -1 &  1 &  1 &  1 &  1 &  1 &  1 &  1 \\
     -1 &  1 & -1 &  1 &  1 & -1 &  1 &  1 &  1 & -1 &  1 & -1 \\
     -1 & -1 &  1 & -1 &  1 &  1 & -1 &  1 &  1 &  1 &  1 &  1 \\
     -1 & -1 & -1 &  1 & -1 &  1 &  1 & -1 &  1 &  1 &  1 &  1 \\
     -1 & -1 & -1 & -1 &  1 & -1 &  1 &  1 &  1 &  1 &  1 &  1 \\
     -1 &  1 & -1 & -1 & -1 &  1 & -1 &  1 &  1 & -1 &  1 & -1 \\
     -1 &  1 &  1 & -1 & -1 & -1 &  1 & -1 &  1 & -1 &  1 & -1 \\
     -1 &  1 &  1 &  1 & -1 & -1 & -1 &  1 &  1 & -1 &  1 & -1
    \end{array}\right].
\end{equation*}
These matrices were constructed using the methods described at the end of Section \ref{sec:resilient_matrix}.

\section{Comparison with Robust Control}\label{apx:robust}

We provide further details of the computation of the ellipsoidal internal bounds \break
$\mathcal{E}\big(x_-(T), X_-(T)\big) \quad \subseteq \quad \mathcal{X}[T]$ on the reach set in Section \ref{subsec:robust}.

The center $x_-(t)$ of each of the internal ellipsoids follows the dynamics
\begin{equation}\label{eq:x_-}
    \dot{x}_- = A x_- + B u_c + C w_c, \quad \text{with} \quad x_-(0) = x_0 \in \mathbb{R}^n,
\end{equation}
and $u_c$ and $w_c$ the respective centers of the control ellipsoid and of the disturbance ellipsoid.

The disturbance ellipsoid is $\mathcal{W} = \mathcal{E}(w_c,Q)$, with its center $w_c := \frac{1}{2}(w_{max} + w_{min})$. The disturbance bounds $w_{min}$ and $w_{max}$ are the mechanical bounds of the uncontrolled actuator defined in \eqref{eq:u}.
We consider loss of control over only one actuator. Thus, $Q$ is a scalar, so $Q(w - w_c)^2 \leq 1$ and $w_{min} \leq w \leq w_{max}$. Hence, $Q = \frac{4}{\left(w_{\max} - w_{\min} \right)^2}$.

\vspace{1mm}
Defining the control ellipsoid is more complicated. To have a fair comparison with the results of our paper, we would need to enforce $\mathcal{L}_2$ bounds on the inputs. However, this is not possible in the framework of \citep{kurzhanski2002}: it allows only for time-invariant ellipsoidal sets of admissible control inputs. Let us find a compromise.
We start from the bounds defined in \eqref{eq:def of U, W, G}: $\|u\|_{\mathcal{L}_2} \leq 1$ and $\|w\|_{\mathcal{L}_2} \leq 1$. So, we want to enforce
\begin{equation*}
    \int_0^T \|u(t)\|^2 dt \leq 1 \qquad \text{and} \qquad \int_0^T \|w(t)\|^2 dt \leq 1,
\end{equation*}
which can be done by choosing $\|u(t)\|^2$, $\|w(t)\|^2 \leq \frac{1}{T}$ for all $t \in [0,T]$. What matters here is the fact that $\|u(t)\|$ and $\|w(t)\|$ have the same bound. Therefore, we choose to limit each input to the smallest of the two intervals $[w_{min}, w_{max}]$ and the interval from \eqref{eq:u}.
The control ellipsoid is then $\mathcal{E}\big(u_c, P\big)$, with its center $u_c := \frac{1}{2}(u_{max} + u_{min})$ and a diagonal shape matrix $P$ with $P_{ii} = \min \left\{ \frac{2^2}{\left(u^i_{\max} - u^i_{\min} \right)^2}, Q \right\}$.

The differential equation for the shape matrix $X_-(t)$ of the internal ellipsoid \citep{kurzhanski2002} is
\begin{equation}\label{eq:X_-}
    \begin{array}{cc}
        \dot{X}_- &= A X_- + X_- A^\top + \sqrt{X_-}S_1(t)B\sqrt{P} + \sqrt{P} B^\top S_1(t) \sqrt{X_-^\top} \\
        &+ \mu\Big( \sqrt{X_-} S_2(t) + S_2(t)\sqrt{X_-^\top}\Big) - \pi(t) X_- - \frac{C Q C^\top}{\sqrt{\pi(t)}},
    \end{array}
\end{equation}
with $X_-(0) = X^0$. The functions $\pi$, $S_1$ and $S_2$ are defined as follows for a given vector $l \in \mathbb{R}^n$:
\begin{equation}\label{eq:terms for X_-}
    \def\arraystretch{1.2}\begin{array}{rlrl}
         l(t) &:= e^{A^\top t} l  \qquad \qquad \qquad \qquad \qquad \pi(t) &:= \sqrt{l(t)^\top C Q C^\top l(t)} \nonumber \\
        S_1(t)B\sqrt{P} &:= \sqrt{\frac{l(t)^\top B P B^\top l(t)}{l(t)^\top X_- l(t)}} \sqrt{X_-} \qquad \qquad S_2(t) &:= \frac{\|l(t)\|}{\sqrt{l(t)^\top X_- l(t)}} \sqrt{X_-}.
    \end{array}
\end{equation}

We can now compute the trajectory of the center of the ellipsoid $x_-(t)$ with \eqref{eq:x_-}, and the evolution of the shape matrix $X_-(t)$ of the ellipsoid with \eqref{eq:X_-}  and \eqref{eq:terms for X_-}. When the radius $\mu$ of the target ball is too small for the target to be reached, then the shape matrix $X_-$ is not positive definite. We investigated for the smallest $\mu$ such that $X_-(T) \succ 0$, and found $\mu = 5.9$. Therefore, the smallest target ball the robust method guarantees to reach has a radius of $5.9$.

\section*{Acknowledgment}

The authors would like to thank Dr. Kenneth Bordignon and Dr. Wayne Durham for providing us the ADMIRE model that enabled our simulations.

% references section

\bibliographystyle{IEEEtran}
\bibliography{biblio.bib}

% biography section

\end{document}